\documentclass[12pt,a4paper]{article}

\usepackage[utf8]{inputenc}
\usepackage[T1]{fontenc}
\usepackage{amsmath,amssymb,amsthm}
\usepackage{mathtools}
\usepackage{xcolor}
\usepackage{multirow}

\usepackage[margin=1in]{geometry}
\usepackage{setspace}
\usepackage{parskip}
\usepackage{titlesec}
\usepackage{enumitem}

\usepackage{graphicx}
\usepackage{float}
\usepackage{subcaption}
\usepackage{tikz}
\usetikzlibrary{arrows.meta,positioning,shapes.geometric,calc}

\usepackage{booktabs}
\usepackage{multirow}
\usepackage{array}
\usepackage{authblk}

\usepackage{algorithm}
\usepackage{algpseudocode}

\usepackage[colorlinks=true,linkcolor=blue,citecolor=blue,urlcolor=blue]{hyperref}
\usepackage[capitalise,noabbrev]{cleveref}

\usepackage{natbib}

\newcommand{\R}{\mathbb{R}}

\newcommand{\argmax}{\operatorname{arg\,max}}

\newcommand{\jk}[1]{\textcolor{black}{#1}}

\newtheorem{definition}{Definition}
\newtheorem{proposition}{Proposition}

\title{%
    \textbf{Adjacent Possible Innovation Dynamics on Local Optima Networks}
}

\author[1,2]{Leonardo Rizzo}
\author[3,4]{Edward D.~Lee}
\author[1]{János Kertész}

\affil[1]{Central European University, Department of Network and Data Science, Vienna, Austria}
\affil[2]{SDA Bocconi, Network Innovation Lab, Milan, Italy}
\affil[3]{Complexity Science Hub, Vienna, Austria}
\affil[4]{University of Natural Resources and Life Sciences, Vienna, Austria}

\date{}

\begin{document}

\maketitle

\begin{abstract}
We propose Local Optima Networks (LONs) as a formal framework for modeling innovation
dynamics. A LON is a directed weighted graph in which nodes represent locally stable
technological configurations and edges encode transition probabilities between their
basins of attraction. We construct LONs from 
fitness landscapes 
and model innovating agents as stochastic walkers \jk{exploring the adjacent possible} on the
resulting network. We show that this 
\jk{model} simultaneously generates the
four main empirical regularities of the discovery-process tradition: sublinear novelty
growth (Heaps' law), heavy-tailed frequency distributions (Zipf's law), anomalous
fluctuation scaling (Taylor's law), and power-law distributed inter-event times. The
exponents fall within empirically observed ranges and are jointly constrained by LON
topology.
Communities in the LON
provide
an operational definition of technological paradigms grounded in basin-level accessibility.
The LON framework thus bridges the discovery-process and adaptive-search traditions of
innovation modeling within a single, parsimonious, and empirically testable representation.

\vspace{1em}
\end{abstract}


\section{Introduction}
\label{sec:introduction}

Formal models of innovation have largely pursued two related but distinct explanatory
tasks. One tradition studies the realized record of novelty as a temporally ordered
sequence and asks why new types appear sublinearly, why reuse is heavy tailed, why
fluctuations scale anomalously, why discoveries arrive in correlated bursts, and why
inter-discovery waiting times follow heavy-tailed distributions. We refer to this as the
\emph{discovery-process} tradition; it is now much broader than a single urn-with-triggering
benchmark: it encompasses species-sampling and Poisson--Dirichlet baselines,
adjacent-possible urns, explicit treatments of Taylor's law, network exploration models
on concept graphs, socially interacting discovery processes, and recent damped or
time-dependent urn schemes designed to accommodate departures from pure power laws
\citep{BlackwellMacQueen1973,PitmanYor1997,Tria2014CorrelatedNovelties,TriaLoretoServedio2018ZipfHeapsTaylor,
Iacopini2018NetworkDynamicsInnovation,marzo2022modeling,AlettiCrimaldi2021Twitter,BellinaEtAl2025}.
A second tradition studies innovation as search over a performance landscape and asks
how interdependent design choices shape path dependence, local trapping, incremental
improvement, and escape to qualitatively new technological approaches. We refer to this as the
\emph{adaptive-search} tradition; it is also richer than a single NK benchmark, examining
decomposability, modularity, hierarchy, cognitive representations, sequencing, distributed
search, competition, and ecosystem interdependence
\citep{dosi1982technological,kauffman1993origins,levinthal1997adaptation,BaumannEtAl2019,frenken2006technological,
Frenken2006Modularity,BillingerEtAl2014,GancoEtAl2020}.

The two traditions 
typically operate at different levels of description. Discovery models explain
sequence-level observables such as novelty growth, revisit frequencies, fluctuation
scaling, and temporal clustering. Adaptive-search models explain how interdependent
choice architectures and search rules shape movement through a performance space. What
is still missing is a 
representation that connects these two levels of
description in a single \jk{framework.}

This paper proposes that \emph{Local Optima Networks} (LONs) can play 
\jk{the bridging role between the two approaches.} A LON
is a directed weighted graph derived from a performance landscape, with nodes representing
local optima and edges representing transition relations between their basins of attraction
\citep{ochoa2008study,verel2011local,ochoa2014lon}. In the combinatorial-optimization
literature, LONs have been used to characterize search difficulty and global landscape
structure, but they remain largely untapped in innovation economics
\citep{khraisha2020complex}. Their appeal here is that
they preserve performance information while also generating a natural state space for
stochastic innovation records.

The LON representation is useful for innovation modeling for three related reasons.
First, it provides a basin-level representation of technological search. In our framework,
a performance landscape encodes the space of configurations and associated performance
values for a given technology. The LON is a coarse-graining of that landscape: its nodes
are local optima, and its edges encode transitions between their basins. Prolonged
residence in the same basin or repeated returns to the same optimum correspond to
incremental exploitation; transitions across basins within the same community correspond
to local reorientation within a technological approach; and transitions across community
boundaries correspond to a shift from one technological paradigm to another. The LON
therefore offers an operational two-level vocabulary for describing innovation dynamics:
basins for incremental innovation, and communities for distinct technological approaches.

Second, the LON links landscape structure to discovery-process observables. A stochastic
walk on a LON generates a time-ordered sequence of visited optima, which can be read as
an innovation record. The statistical properties of that record, including novelty growth,
revisit concentration, fluctuation scaling, and inter-event time distributions, depend on
the topology of the LON: self-loop and residence-time structure affect exploitation
intensity, local connectivity shapes the reachable frontier of new states, and community
structure affects waiting times, burstiness, and temporal clustering. In
\Cref{sec:capabilities}, we show that suitably parameterized walks on LONs generate the
main empirical signatures usually associated with discovery-process models: sublinear
novelty growth (Heaps' law), heavy-tailed reuse (Zipf's law), anomalous fluctuation
scaling (Taylor's law), and power-law distributed inter-event times.

Third, the LON makes structural assumptions explicit and interpretable. In the baseline
model developed below, effective ruggedness is summarized by a persistence parameter
$\rho$ and a small set of additional search parameters. We treat $\rho$ as a reduced-form
proxy for a broader family of structural properties, including ruggedness, basin
persistence, and effective connectivity, rather than as a literal one-to-one substitute for
the full landscape literature. This keeps the model parsimonious while still allowing us
to study how changes in landscape structure reshape the induced LON and, through it, the
realized innovation record.

A final conceptual clarification concerns the unit of novelty. Discovery models often
move between events that are new to a focal process and events that are new to the whole
system. Throughout the paper, it is useful to distinguish \emph{local novelty} (new to
the focal walk, inventor, or firm) from \emph{system-level novelty} (new to the entire
interacting population). The LON framework can accommodate both notions, but the
distinction should be kept explicit.

In terms of contributions, we develop the LON-based model of innovation dynamics
sketched above and show that it generates the main empirical regularities emphasized
by the discovery-process literature while retaining the basin-level,
performance-ordered structure emphasized by the adaptive-search literature. We
characterize, numerically, how LON
structure constrains macroscopic innovation outcomes, including the pace of novelty
growth, the distribution of revisits across states, the emergence of bursty or
heavy-tailed inter-event times, and the fluctuation scaling of innovation counts.

The remainder of the paper is organized as follows. \Cref{sec:litreview} reviews the two
modeling traditions in a way that makes their distinct explanatory targets explicit.
\Cref{sec:model} develops the LON-based innovation model: landscape generation, LON
construction, and walker dynamics. \Cref{sec:lon_properties} characterizes the structural
properties of the resulting LON.
\Cref{sec:capabilities} presents the core results, showing that the model
reproduces the main empirical regularities of the discovery-process tradition within a
framework that retains the structural features of the adaptive-search tradition, and
relates them to the properties of the LON. \Cref{sec:discussion} discusses implications for
innovation theory, possible empirical strategies, and scope conditions.
\Cref{sec:conclusion} concludes.


\section{Two Modeling Traditions \jk{in Innovation Theory} and Their Explanatory Targets}
\label{sec:litreview}

\subsection{The Discovery-Process Tradition}

\subsubsection{Empirical targets}

The discovery-process literature models innovation records as temporally ordered sequences of
tokens drawn from an evolving set of types. Depending on the empirical setting, tokens may be
words, scientific concepts, patents, technologies, songs, hashtags, or other items whose first
appearance can be identified. A useful general distinction is between \emph{local novelty},
meaning new to a focal sequence or agent, and \emph{system-level novelty}, meaning new to the
whole interacting system. This distinction matters in multiagent settings, where something can be
novel for one actor without being globally new \citep{AlettiEtAl2023}.

Four empirical regularities are central in this tradition.

\paragraph{Heaps' law (novelty growth).}
Let $D(t)$ denote the number of distinct items observed after $t$ draws. In many empirical
systems, one finds
\begin{equation}
    D(t) \sim t^{\beta}, \qquad 0 < \beta < 1,
\end{equation}
so that the number of distinct observed types grows sublinearly with sequence length. The
interpretation is that novelty continues to appear, but at a decelerating rate. Sublinear growth
is often treated as the baseline empirical signature of innovation in temporally ordered records
\citep{Tria2014CorrelatedNovelties,TriaLoretoServedio2018ZipfHeapsTaylor}.

\paragraph{Zipf's law (reuse concentration).}
Let $f(R)$ denote the frequency of a
\jk{basin of a technology having} rank $R$. In many innovation-related data sets,
the frequency--rank relation is heavy tailed and is often approximated by
\begin{equation}
    f(R) \sim R^{-\alpha}.
\end{equation}
The exponent $\alpha$ summarizes how concentrated reuse is among previously discovered
elements. In idealized cases, the Heaps and Zipf exponents satisfy the reciprocal relation
$\alpha \approx 1/\beta$, but empirical deviations are common and can reflect correlations,
nonstationarity, and finite-size effects \citep{Tria2014CorrelatedNovelties,TriaLoretoServedio2018ZipfHeapsTaylor}.

\paragraph{Taylor's law (fluctuation scaling).}
Let $n_{\Delta}(t)$ be the number of innovation events in a time window $[t,t+\Delta)$. Across
many systems, the variance of $n_{\Delta}$ scales as a power of its mean:
\begin{equation}
    \mathrm{Var}(n_{\Delta}) \sim \mathbb{E}[n_{\Delta}]^{\gamma}.
\end{equation}
The benchmark $\gamma = 1$ corresponds to
Poisson-like fluctuations, while $\gamma > 1$ indicates clustered activity and stronger-than-Poisson variability. Recent work on innovation processes argues that Taylor's law should be
treated not as an optional add-on to Heaps and Zipf, but as a coequal empirical constraint on
any adequate generative model \citep{TriaLoretoServedio2018ZipfHeapsTaylor,TriaCrimaldiAlettiServedio2020TaylorInnovation}.

\paragraph{Inter-event time distribution (burstiness).}
Let $\tau$ denote the time between consecutive novelty events, that is, the inter-event
or inter-discovery time. In many empirical systems, the distribution of $\tau$ follows a
heavy-tailed power law:
\begin{equation}
    p(\tau) \sim \tau^{-\gamma_{\mathrm{IET}}}.
    \label{eq:iet}
\end{equation}
A memoryless (Poisson) discovery process would produce exponentially distributed
inter-event times; the empirical observation of power-law tails indicates that discoveries are bursty, arriving in
temporally clustered episodes separated by long quiescent intervals
\citep{BarabasiBurstiness2005,Iacopini2018NetworkDynamicsInnovation}. Burstiness is closely related to
Taylor's law: both reflect the same underlying departure from Poisson regularity, but
the inter-event time distribution characterizes the temporal fine structure of discovery
at the level of individual waiting times rather than windowed counts.

Taken together, these four regularities imply that the discovery-process tradition is concerned
not only with how often novelty appears, but also with how novelty, reuse, fluctuations,
and temporal clustering are jointly organized over time.

\subsubsection{Mechanistic families}

The discovery-process literature can be organized as a sequence of increasingly structured
model families. The simplest starting point is the species-sampling family, including
Blackwell--MacQueen, Dirichlet, and Poisson--Dirichlet schemes, in which the probability
of seeing an old type is proportional to how often it has appeared before and the probability
of a new type is governed by an innovation term. These processes provide an analytically
tractable baseline for preferential reuse and sublinear novelty growth, but the space of
possibilities is not itself history dependent
\citep{BlackwellMacQueen1973,PitmanYor1997}.

A major step beyond exchangeable species-sampling models is to let novelty expand the space
of future novelties. This is the core idea of the adjacent possible, formalized in the urn model
with triggering and related variants, where the appearance of a novelty introduces further
latent possibilities and the opportunity set becomes endogenous to past discoveries. This class
of models can generate Heaps' and Zipf's laws within a common mechanism and, in later work,
Taylor's law as well \citep{Tria2014CorrelatedNovelties,TriaLoretoServedio2018ZipfHeapsTaylor}. Discovery models need not
treat the possibility space as an unstructured urn, however. An alternative is to represent
concepts as nodes of a graph and model discovery as a reinforced walk, so that semantic
locality and temporal correlations emerge because walkers preferentially reuse previously
traversed edges while still discovering new nodes
\citep{Iacopini2018NetworkDynamicsInnovation}.

More recent work makes the discovery process explicitly social: multiple agents explore a space
of possibilities while interacting on a social network, so that one agent's discoveries affect the
future discoveries of others. The adjacent possible thereby acquires a social dimension alongside
its semantic one, since an actor's network position shapes the ideas or objects to which they
become exposed \citep{IacopiniEtAl2020}. A further development is the move
away from strictly stationary power-law urns. Damped innovation models modify the update
function for old items to fit curved rank--frequency plots and different discovery regimes, while
time-dependent urns allow reinforcement and triggering to vary over time or with the number
of discovered types, broadening the tradition from canonical power laws to a more general
account of how innovation records bend, cross over, or saturate
\citep{AlettiCrimaldi2021Twitter,BellinaEtAl2025}. The newest strand explicitly models several
innovation processes that mutually influence one another, distinguishing node-level discoveries
from system-level novelties and developing asymptotic and inferential tools for estimating
interaction strengths \citep{AlettiEtAl2023,AlettiEtAl2025}.

Even in its richer networked and interacting forms, the discovery-process tradition primarily
models the appearance, recurrence, and correlation of novelties. It rarely endogenizes a
performance ordering over states, the basin geometry around locally superior solutions, or the
barrier structure that separates incremental improvement from genuine escape. Those are the
issues on which adaptive-search models remain especially informative.

\subsection{The Adaptive-Search Tradition}

\subsubsection{Landscape structure and search}

The adaptive-search tradition represents innovation as movement on a performance landscape.
A configuration $s$ in a design space $\mathcal{S}$ is associated with a performance value
$f(s)$, and search consists of moving through $\mathcal{S}$ by proposing and evaluating
alternative configurations. This approach goes back to the landscape metaphor in evolutionary
theory and entered organization and innovation research through NK-style models of
interdependent choices \citep{wright1932roles,kauffman1993origins,levinthal1997adaptation}.

In the canonical NK model, ruggedness increases with the degree of interdependence among
choices, making local hill climbing more likely to end at a suboptimal peak. But the
literature has long moved beyond a single scalar view of complexity. Review work now
treats rugged search as a family of problems that vary in decomposability, modularity,
hierarchy, cognitive representability, and the possibility of distributed or sequential search
\citep{BaumannEtAl2019,frenken2006technological}. Several structural features recur across
this work. Interdependence generates multiple peaks and valleys, making myopic search path
dependent; basins of attraction determine where local search is likely to end up, while the
distribution of local optima shapes the trade-off between reliable local improvement and the
risk of getting trapped \citep{levinthal1997adaptation,kauffman1993origins}. The difficulty of search depends
not only on how many interdependencies there are, but on how they are organized: near-decomposable
or modular architectures permit partial problem decomposition, whereas hierarchical or asymmetric
structures make some choices more influential than others
\citep{Frenken2006Modularity,ethiraj2004modularity,BaumannEtAl2019}. It is also important to note that the path
dependence generated by fitness-landscape models, where local search trajectories depend on
initial conditions and the sequence of moves on a fixed landscape, is analytically distinct from
the market lock-in studied in increasing-returns adoption models, where payoffs themselves
change through cumulative adoption \citep{Frenken2006Modularity,frenken2006technological}.

\subsubsection{Search processes and organizational settings}

The adaptive-search literature is not only about landscape structure; it is also about how search
is organized. A central issue is whether search is local and incremental or involves long jumps
that change multiple choices at once: broader search can help escape low peaks, but it can also
waste resources or abandon promising improvements too early
\citep{BaumannEtAl2019}. Experimental work further shows that search behavior adapts to
performance feedback, with success narrowing search to the local neighborhood and repeated
failure inducing more exploratory moves \citep{BillingerEtAl2014}. Search is often distributed
across multiple actors, whether organizational units solving interdependent subproblems or
upstream and downstream firms innovating in the same ecosystem; the structure of technological
interdependence and input--output flows jointly shape the returns to narrow optimization versus
mixing and matching \citep{GancoEtAl2020}. Several extensions abandon the assumption of a
fixed landscape entirely: dynamic NK models with environmental turbulence show that broader
search becomes relatively more valuable when the environment changes rapidly or extensively,
bringing the adaptive-search tradition closer to the temporal concerns of discovery-process
models \citep{LiChenYing2019}.

Adaptive-search models richly describe how landscape structure and search rules affect
performance trajectories, but they do not usually take the empirical regularities of realized
innovation records as primary targets. They typically measure adaptation by performance
improvement, reached peaks, or time to convergence rather than by the statistics of the
resulting innovation sequence.

\subsection{Local Optima Networks as a Family of Bridge Representations}
\label{subsec:lit_lons}

Local Optima Networks (LONs) compress a fitness landscape into a directed,
weighted graph whose nodes are local optima and whose edges represent
transition relations between them \citep{ochoa2008study,ochoa2014lon}.
Foundational work defined two main edge semantics. In \emph{basin-transition}
LONs, edge weights approximate the probability that a move from one basin of
attraction reaches another basin. In \emph{escape-edge} LONs, edges record the
probability that a bounded perturbation followed by local search escapes one
optimum and reaches another \citep{verel2011local}. These constructions already
showed that LON metrics such as the number of local optima, path length to the
global optimum, clustering, modularity, and weight disparity are informative
about search difficulty and can correlate with heuristic performance
\citep{chicano2012autocorrelation,ochoa2014lon}.

A key lesson from later work is that there is no single canonical LON. Different
construction procedures preserve different aspects of the landscape. Markov-chain
sampling records adaptive walks on the local-optima level, whereas snowball
sampling recursively explores local-optima neighborhoods from a branching
random walk. These two samplers can produce quite different network summaries
of the same problem instance, and their features can be complementary rather than
substitutable \citep{ThomsonEtAl2020SamplingLON}. In that sense, a LON should be
understood not as one graph but as a family of coarse-grainings defined by a
choice of nodes, edges, and sampling rule.

This point is reinforced by method-induced LONs. In memetic differential
evolution, for example, parent-child relations between local optima discovered
during the run can themselves define a growing weighted network, and network
centrality can then be used online to guide future parent selection
\citep{HomolyaVinko2020LeveragingLONMDE,HomolyaVinko2019CentralityMDE}. Such constructions do not only
summarize a landscape; they summarize how a particular heuristic actually
discovers that landscape. This distinction between structural LONs and
trace-induced LONs is useful for innovation modeling, where observed innovation
records may reveal the latter more readily than the former.

Recent applications also show that neutrality and representation choice can be
substantive. In parameter-landscape analysis, monotonic and compressed monotonic
LONs collapse equal-fitness plateaus and distinguish optimal from suboptimal
sinks, revealing that parameter spaces may be much more multimodal than
previous slice-based analyses suggested \citep{CleghornOchoa2021PSOLON}. In feature
selection, compressed monotonic LONs identify neutral plateaus among global
optima, and these plateaus can reveal irrelevant features directly
\citep{MostertEtAl2019FeatureSelectionLON}. In AutoML search spaces, varying neighborhood size
and perturbation radius changes the number of basins, sinks, and effective
distances between optima, showing that search difficulty depends strongly on
how local reachability is defined \citep{TeixeiraPappa2022AutoMLLON}. In morpho-evolution,
different encodings of the same task induce markedly different LONs: some
show self-loops and short chains, while others generate longer improving
trajectories and greater ability to escape poor local optima
\citep{ThomsonEtAl2024MorphoLON}.

These developments are especially important for innovation theory. They imply
that a LON is not merely a generic network layered on top of innovation data.
It is a coarse-graining of the performance landscape itself, but one whose exact
form depends on substantive modeling choices. This makes LONs attractive as a
bridge between adaptive-search models and discovery-process models. On the
one hand, they retain information about basins, barriers, sinks, neutrality,
community structure, and escape. 
A stochastic process on a LON
generates an innovation record whose novelty growth, revisit concentration,
waiting-time structure, and burstiness can be studied directly.

For the purposes of this paper, one concept is particularly useful: \emph{navigability}.
By navigability we mean the ease with which search both escapes local optima and
continues to discover distinct, high-performing regions of the state space. This
extends the usual focus on ruggedness alone and gives a more precise language for
relating LON structure to observed innovation dynamics.


\section{The LON-Innovation Model}
\label{sec:model}

Our model consists of three integrated components: (1) a toroidal \jk{(periodic boundary condition)} fitness landscape
representing the space of technological configurations and their performance levels; (2)
a Local Optima Network extracted from the landscape and representing its macroscopic
structure; and (3) a walker dynamics model representing how innovating agents navigate
the LON over time.

\subsection{Fitness Landscape Generation}
\label{subsec:landscape}

\subsubsection{Conceptual framework}

A fitness landscape is a mapping from a configuration space to a performance measure.
In the innovation context, configurations represent technological designs, organizational
routines, or strategic positions; performance measures profitability, efficiency, or
market share.

We work with a discrete two-dimensional configuration space
$\mathcal{S} = \{0, 1, \ldots, L-1\}^2$. The two-dimensional case preserves all key
qualitative features (multiple optima, basins, communities, ruggedness variation) while
enabling visualization and tractable analysis. Extensions to higher dimensions are
discussed in \Cref{sec:discussion}.

\begin{definition}[Fitness Landscape]
\label{def:landscape}
A fitness landscape is a tuple $\mathcal{L} = (\mathcal{S}, f, \mathcal{N})$ where:
$\mathcal{S}$ is the configuration space; $f: \mathcal{S} \rightarrow \R$ is the fitness
function; and $\mathcal{N}: \mathcal{S} \rightarrow 2^{\mathcal{S}}$ is the neighborhood
structure defining which configurations are reachable via incremental innovation.
\end{definition}

\subsubsection{Toroidal topology}

A critical design choice concerns the boundary conditions of the landscape. Standard
reflecting or absorbing boundaries introduce artifacts: artificial barriers or traps at
the edges. We impose instead \emph{periodic} (toroidal) boundary conditions:

\begin{definition}[Toroidal Configuration Space]
\label{def:torus}
The toroidal configuration space $\mathcal{S}_T$ is defined such that
$(x + L, y) \equiv (x, y)$ and $(x, y + L) \equiv (x, y)$ for all $(x, y)$. The
toroidal distance between two points is:
\begin{equation}
    d_T\big((x_1, y_1), (x_2, y_2)\big) = \sqrt{\min(|x_1 - x_2|, L - |x_1 - x_2|)^2
    + \min(|y_1 - y_2|, L - |y_1 - y_2|)^2}
\end{equation}
\end{definition}

The toroidal structure represents an innovation environment without privileged positions:
no configuration is inherently central or peripheral, consistent with the abstract,
theory-building purpose of the model.

\subsubsection{Perlin noise generation via the Clifford torus}

We generate the fitness function $f$ using 
\jk{fractional } Brownian motion (fBm) built from
four-dimensional Perlin noise \citep{perlin1985image}. The key technical challenge
is ensuring that the noise tiles seamlessly on the torus. We solve this by mapping
the two-dimensional grid onto a four-dimensional Clifford torus:

\begin{definition}[Clifford Torus Mapping]
\label{def:clifford}
The mapping $\phi: [0, 1)^2 \rightarrow \R^4$ is:
\begin{equation}
    \phi(u, v) = \left( \cos(2\pi u),\ \sin(2\pi u),\ \cos(2\pi v),\ \sin(2\pi v) \right)
\end{equation}
where $(u, v) = (x/L, y/L)$. Adjacent points on the 2D grid (including across boundaries)
map to adjacent points in 4D space, guaranteeing seamless tiling.
\end{definition}

\begin{definition}[fBm Fitness Function]
\label{def:fbm}
The fitness function with $n$ octaves is:
\begin{equation}
    f(x, y) = \sum_{i=0}^{n-1} \rho^i \cdot \eta\left( \phi\left(\frac{x}{L},
    \frac{y}{L}\right) \cdot \lambda^i \cdot \omega \right)
\end{equation}
where $\eta$ is a 4D Perlin noise function; $\omega$ is the base frequency; $\lambda$
is the lacunarity (frequency multiplier per octave); $\rho \in (0,1)$ is the
\emph{persistence} (amplitude multiplier per octave); and $n$ is the number of octaves.
\end{definition}

The persistence parameter $\rho$ is the primary control for landscape ruggedness. Low
$\rho$ down-weights high-frequency octaves, producing smooth landscapes with few local
optima and large basins. High $\rho$ gives high-frequency components near-equal weight,
producing rugged landscapes with many local optima and small basins. This provides a
continuous analog to the epistasis parameter $K$ in NK landscapes, with a direct
interpretation: $\rho$ controls the degree to which fine-grained technological details
interact with coarse-grained design choices.

The final landscape is normalized to $[0, 100]$:
\begin{equation}
    F(x, y) = 100 \cdot \frac{f(x, y) - f_{\min}}{f_{\max} - f_{\min}}
\end{equation}


\subsection{Local Optima and Basins of Attraction}
\label{subsec:optima}

\subsubsection{Neighborhood and local optima}

The neighborhood structure $\mathcal{N}$ defines the set of configurations reachable
from any given point by a single incremental innovation step. We use the
\emph{von Neumann} (4-connected) neighborhood, where the neighbors of
$(x, y)$ are the four cardinal-direction points $(x \pm 1, y)$ and $(x, y \pm 1)$,
representing constrained innovation where only one attribute of the configuration is
modified at a time. The neighborhood is computed modulo $L$ to respect toroidal boundary
conditions.

\begin{definition}[Local Optimum]
\label{def:local_optimum}
A configuration $s^* \in \mathcal{S}$ is a local optimum if $f(s^*) \geq f(s)$ for all
$s \in \mathcal{N}(s^*)$. The global optimum is the local optimum with highest fitness.
\end{definition}

Local optima represent stable technological configurations from which no incremental
modification improves performance. An agent that has converged to such a configuration
via incremental search exhibits technological lock-in: it cannot improve through any
marginal adjustment available within its neighborhood.

\subsubsection{Hill climbing and optimum identification}

Local optima are identified via \emph{steepest-ascent hill climbing}, a deterministic
local search procedure that also defines the basin structure of the landscape.

\begin{definition}[Steepest-Ascent Hill Climbing]
\label{def:hill_climb}
Starting from an initial configuration $s_0 \in \mathcal{S}$, the hill climbing
procedure generates a sequence $s_0, s_1, s_2, \ldots$ by the update rule:
\begin{equation}
    s_{t+1} = \argmax_{s \in \{s_t\} \cup \mathcal{N}(s_t)} f(s)
\end{equation}
The sequence terminates at the first $t^*$ such that $s_{t^*+1} = s_{t^*}$, i.e.,
when no neighbor improves upon the current configuration. The terminal state
$s^* = s_{t^*}$ is a local optimum.
\end{definition}

Each step selects the single best-improving neighbor, making the ascent greedy and
deterministic. The toroidal topology ensures that the neighborhood is well-defined
everywhere, with no boundary artifacts. The procedure always terminates in finite time
because fitness is bounded and strictly increases at each non-terminal step.

Steepest-ascent hill climbing, as opposed to first-improvement variants, is the natural
choice here because it is fully deterministic: every starting point maps to exactly one
terminal optimum, making the basin of attraction well-defined and uniquely computable
without stochastic runs. This is essential for the LON construction in
\Cref{subsec:lon}, which requires a clean partition of the configuration space into
non-overlapping basins.

To identify the complete set of local optima in a landscape of size $L \times L$, we
apply hill climbing exhaustively from every grid point. Each starting location yields a
terminal optimum; duplicate results, that is, pairs of identified optima within a small
positional tolerance $\epsilon$ of one another, to account for gradient ties on the
discrete grid, are merged by retaining the higher-fitness representative. The final
result is a de-duplicated list of local optima $\{s^*_1, \ldots, s^*_n\}$ sorted by
decreasing fitness.

\subsubsection{Basins of attraction}

\begin{definition}[Basin of Attraction]
\label{def:basin}
The basin of attraction $B(s^*)$ of a local optimum $s^*$ is the set of all
configurations from which steepest-ascent hill climbing converges to $s^*$:
\begin{equation}
    B(s^*) = \{s \in \mathcal{S} : \mathrm{HillClimb}(s) = s^*\}
\end{equation}
The basin size $|B(s^*)|$ measures the number of configurations in the space from which
$s^*$ is reachable by incremental improvement.
\end{definition}

Because each grid point converges to exactly one optimum, the basins partition
$\mathcal{S}$ exhaustively and without overlap:
\begin{equation}
    \bigcup_{i=1}^{n} B(s^*_i) = \mathcal{S}, \qquad B(s^*_i) \cap B(s^*_j) =
    \emptyset \text{ for } i \neq j
\end{equation}

The basin size distribution is a key structural property of the landscape.
It governs the probability that an agent initiating random exploration will
converge to a given optimum, and therefore shapes the likelihood distribution
over which technology the agent will discover. Large basins represent robust,
easily discoverable technologies; small basins represent fragile configurations
that require precisely directed search to reach.

Basin size also correlates with the self-loop structure of the LON, as formalized
in \Cref{subsec:lon}: a larger basin implies that random perturbations from the optimum
are more likely to remain within the same basin and return to the same optimum via hill
climbing, yielding a higher self-loop weight $w_{ii}$ and thus stronger path dependence
at that configuration.

\subsubsection{Base frequency and the $\omega L$ ratio}
\label{subsubsec:omegaL}

The fitness function is generated via fBm in which the lowest-frequency octave oscillates
at frequency $\omega$. The natural dimensionless control is the product $\omega L$, which
counts the number of base-frequency cycles that fit within the grid and therefore
determines how many dominant performance peaks are present in the configuration space.
Large $\omega L$ (e.g., $\omega L = 600$) produces a rich mosaic of hills and valleys
with many local optima; small $\omega L$ (e.g., $\omega L = 120$) yields a landscape
dominated by one or two broad features. Holding $\omega L$ fixed while scaling both
$\omega$ and $L$ proportionally changes only the resolution, not the qualitative
structure, so we treat $\omega L$ as the primary control for macro-scale landscape
richness.

\begin{figure}[H]
    \centering
    \includegraphics[width=\textwidth]{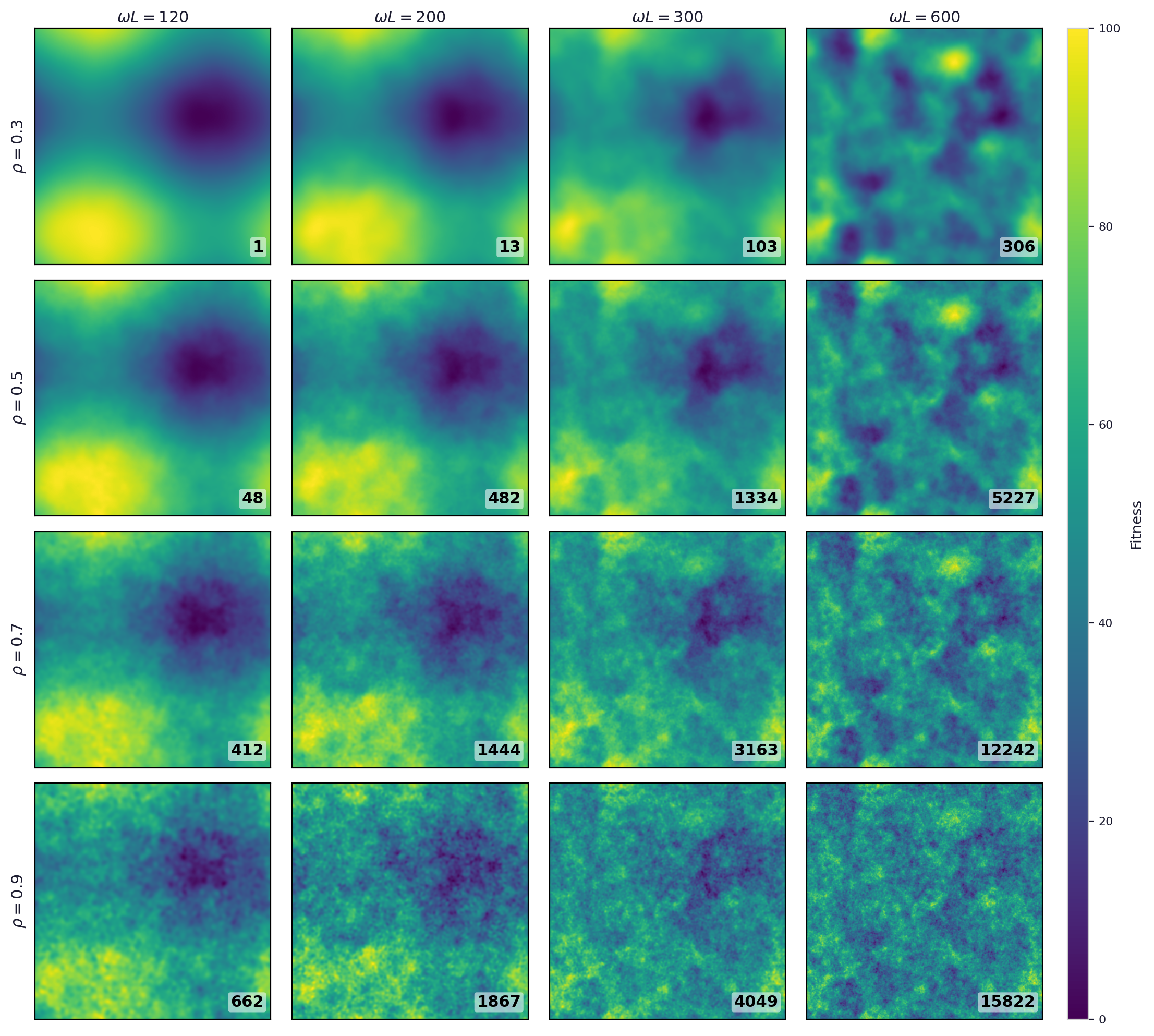}
    \caption{Fitness landscapes varying $\omega L$ (columns, increasing left to right)
    and persistence $\rho$ (rows, increasing top to bottom), with $L = 1000$ and
    $n = 6$ octaves. The number of local optima identified by exhaustive hill climbing
    is shown in each panel. Fitness is normalized to $[0, 100]$.}
    \label{fig:landscape_grid}
\end{figure}

The distinction between $\omega L$ and $\rho$ maps onto a substantively meaningful
separation: $\omega L$ controls how many broad, distinct high-performing approaches
exist, while $\rho$ controls the fine-grained ruggedness within each approach. These
two levels are independently calibrated, allowing the model to represent a technology
with few broad approaches but many internal traps ($\omega L$ small, $\rho$ high), or
many competing approaches that are each individually smooth ($\omega L$ large,
$\rho$ low). \Cref{fig:landscape_grid} illustrates the joint effect across
$4 \times 4 = 16$ representative landscapes.

\subsection{Local Optima Network Construction}
\label{subsec:lon}

\subsubsection{Network definition}

\begin{definition}[Local Optima Network]
\label{def:lon}
A Local Optima Network is a directed weighted graph $G = (V, E, w)$ where nodes
$V = \{v_1, \ldots, v_n\}$ represent local optima; edges $E \subseteq V \times V$
represent possible transitions between basins; and weights $w: E \rightarrow [0,1]$
are transition probabilities. Each node $v_i$ carries attributes: position $(x_i, y_i)$,
fitness $f_i = F(x_i, y_i)$, and basin size $|B(v_i)|$.
\end{definition}

\begin{figure}[H]
    \centering
    \includegraphics[width=\textwidth]{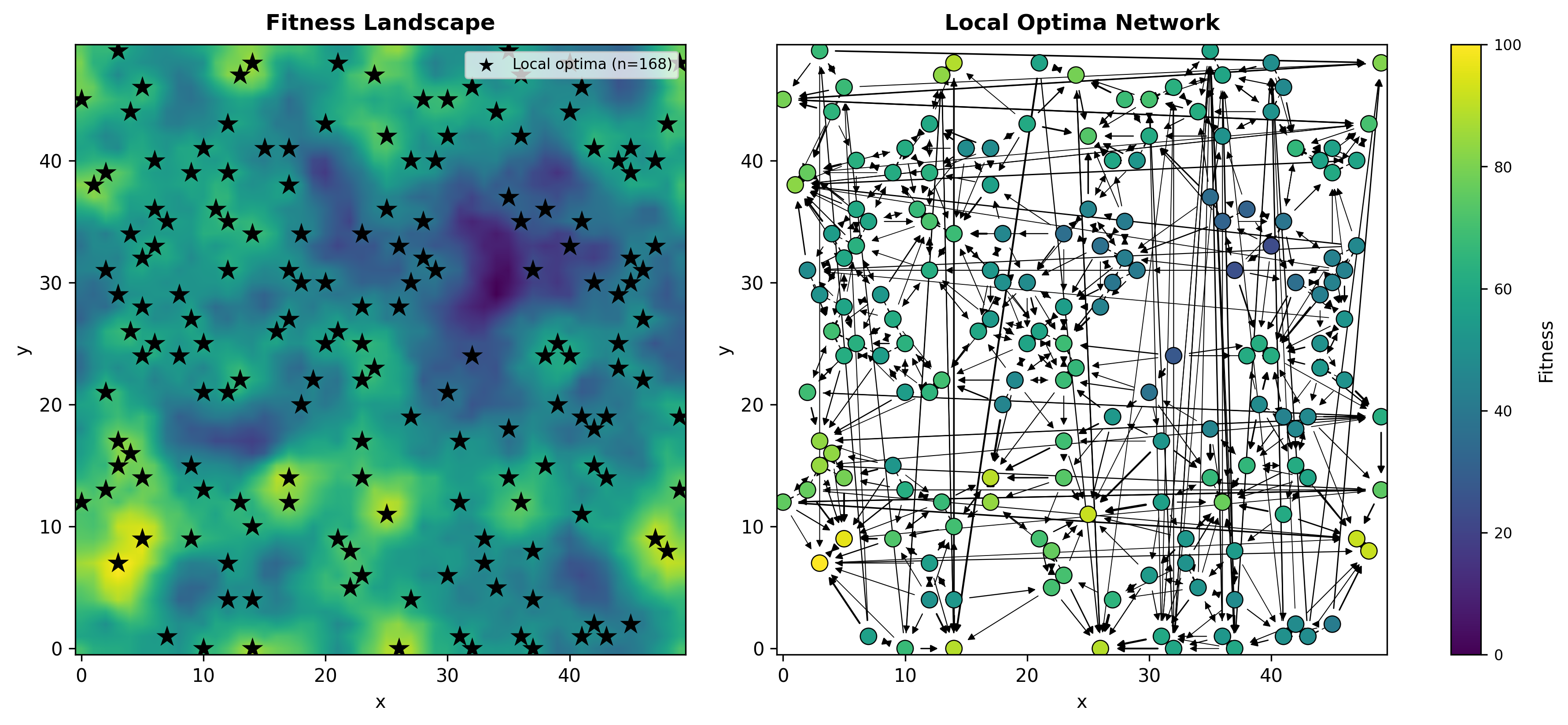}
    \caption{Left: a representative fitness landscape with local optima marked as
    stars. Right: the corresponding Local Optima Network, with nodes positioned at
    their landscape coordinates and colored by fitness $f_i$ (same colormap).
    Directed edges represent inter-basin transitions with probability $w_{ij} > 0$;
    self-loops ($w_{ii}$) are omitted for clarity. Edges that cross the boundary of
    the panel connect nodes that are spatially adjacent on the torus.}
    \label{fig:lon_example}
\end{figure}

\Cref{fig:lon_example} illustrates the relationship between a fitness landscape and
its corresponding LON for a representative parameter configuration. The left panel
shows the landscape with its local optima marked as stars; the right panel shows the
resulting LON, where nodes are positioned at their spatial coordinates in the
landscape, node color encodes fitness, and directed edges encode inter-basin
transitions. Self-loop edges, that is, transitions from a node back to itself
representing perturbations that remain within the same basin, are not shown; for many
nodes, particularly those sitting atop the broad high-fitness hills, the self-loop
weight $w_{ii}$ is the dominant outgoing weight. Several edges span what appears to be
the full width or height of the panel, connecting nodes on opposite sides of the grid;
these are not anomalous long-range connections but a direct consequence of the toroidal
topology, in which the left and right borders are identified and the same holds for top
and bottom.

\subsubsection{\jk{Weighted} edge construction via basin-hopping}

We construct edges using a \emph{basin-hopping} procedure. From each optimum $v_i$,
we sample $m$ random 
\jk{initial points} uniformly from the ball of radius $r$ centered at
$v_i$, apply hill climbing to each of them,
and record which basin is reached.
The edge weight $w_{ij}$ is the fraction of 
\jk{points around} $v_i$ that converge
to $v_j$:

\begin{equation}
    w_{ij} = \frac{1}{m} \sum_{k=1}^{m} \mathbf{1}[\mathrm{HillClimb}(v_i + \delta_k) = v_j],
    \qquad \delta_k \sim \mathrm{Uniform}([-r, r]^2)
\end{equation}

Weights are normalized so that outgoing edges from each node sum to one:
\begin{equation}
    \sum_{j : (i,j) \in E} w_{ij} = 1 \quad \forall i \in V
\end{equation}

The weight $w_{ij}$ has a direct interpretation: it is the probability that an agent
at local optimum $i$, upon experiencing a random perturbation within radius $r$, will be
directed to optimum $j$ by subsequent hill climbing. The self-loop weight $w_{ii}$
measures basin robustness, the probability that the agent returns to its current
configuration after perturbation, while the escape probability $1 - w_{ii}$ measures
the ease of transitioning to a different locally stable technological approach.

\subsubsection{Perturbation radius and the characteristic inter-optimum distance}
\label{subsubsec:perturbation_radius}

The perturbation radius $r$ is the most consequential structural parameter of the LON
construction. It determines not only the weights on individual edges but the global
connectivity of the network. To set $r$ in a principled way, we define the
\emph{characteristic inter-optimum distance}:
\begin{equation}
    d^* = \frac{L}{\sqrt{n}}
    \label{eq:dstar}
\end{equation}
where $n$ is the number of local optima. This quantity is not a free parameter but an
emergent property of the landscape: since $n$ is determined by $\omega L$ and $\rho$,
$d^*$ decreases as landscapes become more rugged and increases as they become smoother.

The ratio $r / d^*$ governs a percolation transition in the LON's connectivity.\footnote{According to our experience and the parameter space that we explored, the percolation transition happens at around $r / d^* = 0.25$} When
$r \ll d^*$, perturbations from any optimum are almost entirely confined within its own
basin: the LON consists of isolated self-loops with negligible inter-node edges. When $r \gg d^*$, the LON is densely connected and the
fitness-correlated heterogeneity that gives the network its innovation-theoretic content
is progressively destroyed.

In the baseline model, we fix $r \approx d^*$. This is the regime of richest structure and most sensitive dynamics: the
network is globally connected but sparse, with high-fitness broad-basin nodes retaining
high self-loop weights while low-fitness narrow-basin nodes provide the connective tissue.
Increasing $r$ above $d^*$ erodes the self-loop dominance of even the highest-fitness
nodes and broadens the out-degree distribution, moving the walk toward uniform mixing;
decreasing $r$ below $d^*$ fragments the network into isolated components, trapping
agents at their initial optima.

\subsection{Walker Dynamics Model}
\label{subsec:walker}

\subsubsection{The walker as an innovating agent}

An innovating agent, whether a firm, research team, or inventor, is modeled as a
stochastic walker on the LON. The walker's state $v_t \in V$ at time $t$ is the local
optimum it currently occupies.

\begin{definition}[LON Walker]
\label{def:walker}
A walker on LON $G = (V, E, w)$ is characterized by its current node $v_t$ and the
Markov transition rule:
\begin{equation}
    \Pr(v_{t+1} = j \mid v_t = i) = w_{ij}
\end{equation}
The walker's fitness history $\{f_{v_0}, f_{v_1}, \ldots\}$ and novelty record
$\{t : v_t \notin \{v_0, \ldots, v_{t-1}\}\}$ are the primary observables.
\end{definition}

Each step represents a period of incremental exploration, after which the agent either
returns to the same configuration (
\jk{reflecting the basin's robustness}) or transitions to a
neighboring basin (if the perturbation exceeds the basin boundary). The self-loop
weight $w_{ii}$ directly controls the probability of remaining in the current
configuration, operationalizing path dependence.

\subsubsection{Innovation record and novelty events}

From the walker's trajectory, we define two key observables that connect to the
discovery process literature:

\begin{definition}[Novelty Rate and Revisit Rate]
\label{def:novelty}
Let $\mathcal{V}(t) = \{v_0, v_1, \ldots, v_t\}$ be the set of distinct optima visited
by time $t$. The \emph{novelty count} is $D(t) = |\mathcal{V}(t)|$. The \emph{visit
frequency} of optimum $v$ is $n_v(t) = |\{\tau \leq t : v_\tau = v\}|$.
\end{definition}

The novelty count $D(t)$ is the direct analog of the distinct-item count in urn models.
Its growth rate determines whether the walk exhibits Heaps-like behavior. The visit
frequency distribution $\{n_v(t)\}_{v \in \mathcal{V}(t)}$ is the analog of the
rank-frequency distribution in Zipf's law. The sequence of times at which new optima
are first visited defines the inter-event time series whose distribution is the analog
of the burstiness signature in \Cref{eq:iet}.

\subsubsection{Exogenous innovation: A further escape mechanism}
\label{subsubsec:escape}

The baseline model \jk{as described above }
is driven entirely by the
basin-hopping dynamics encoded in the transition weights $w_{ij}$, \jk{i.e., by endogenous innovations.} As shown in
\Cref{sec:capabilities}, this is sufficient to generate all four empirical
regularities. The model can, however, be augmented with a small probability
$\varepsilon > 0$ of teleporting to a uniformly random node at each step:
\begin{equation}
    \Pr(v_{t+1} = j \mid v_t = i) =
    (1 - \varepsilon)\, w_{ij} \;+\; \frac{\varepsilon}{n}
    \label{eq:epsilon_transition}
\end{equation}
This teleportation represents a rare exogenous event, such as an external technology
transfer or a recombination insight, that places the agent in a region of the
configuration space unreachable by incremental search. For any $\varepsilon > 0$,
\Cref{eq:epsilon_transition} defines an ergodic Markov chain with a unique stationary
distribution over all $n$ nodes. The effects of introducing a small $\varepsilon$ on
the four innovation exponents are explored in \Cref{subsec:param_variations}.

\subsection{Model Parameters}
\label{subsec:params}
 
\Cref{tab:all_params} provides a complete summary of the model's parameters,
organized by the component in which they appear. The landscape is characterized
by the base frequency $\omega$ and grid size $L$, whose product $\omega L$
controls the macro-scale richness of the performance surface together with
persistence $\rho$, which governs fine-scale ruggedness. The LON construction
is determined by the neighborhood type $\mathcal{N}$ and perturbation radius $r$,
the latter fixed at the characteristic inter-optimum distance
$d^* = L/\sqrt{n}$. Walker dynamics are fully determined by the basin-hopping
transition weights in the baseline ($\varepsilon = 0$); the optional exogenous
innovation rate $\varepsilon$ can be set to a small positive value to introduce
rare random jumps.
 
\begin{table}[H]
\centering
\caption{Model parameter summary.}
\label{tab:all_params}
\begin{tabular}{@{}p{2.8cm}p{3.2cm}p{1.2cm}p{6.2cm}@{}}
\toprule
\textbf{Component} & \textbf{Parameter} & \textbf{Symbol}
    & \textbf{Interpretation} \\
\midrule
\multirow{4}{*}{\rotatebox[origin=c]{90}{\parbox{3.3cm} {Landscape}}}
    & Grid size        & $L$      & Resolution of the configuration space \\
    & Base frequency   & $\omega$ & Spatial scale of dominant features;
                                    $\omega L$ counts feature cycles across the grid \\
    & Persistence      & $\rho$   & Fine-scale ruggedness; high $\rho$ yields
                                    many optima with small basins \\
    & Octaves          & $n$      & Number of superimposed frequency layers \\
\midrule
\multirow{3}{*}{\rotatebox[origin=c]{90}{{\parbox{2.5cm} {LON \\ construction}}}}
    & Neighborhood     & $\mathcal{N}$ & Incremental move set\ 
                                         (von Neumann, 4-connected) \\
    & Perturbation radius & $r$   & Reach of non-incremental search;
                                    fixed at $d^* = L/\sqrt{n}$ \\
    & Perturbation samples & $m$  & Number of random draws per node
                                    for edge weight estimation \\
\midrule
\multirow{1}{*}{\rotatebox[origin=c]{90}{\parbox{1.4cm} {Walker}}}
    & Exogenous innovation & $\varepsilon$ & Probability of teleporting to
                                             a uniformly random node at each step;
                                             $\varepsilon = 0$ in the baseline \\
\bottomrule
\end{tabular}
\end{table}


\section{Structural Properties of the Local Optima Network}
\label{sec:lon_properties}

The LON constructed in \Cref{subsec:lon} is not merely a convenient summary of the
landscape: its topology encodes the full set of constraints on how an innovating agent
can move through the space of locally stable technological approaches. Before turning
to walker dynamics and the resulting innovation records, it is worth characterizing the
structural properties that emerge from the construction itself.

\subsection{Fitness, Self-Loop Weight, and Incoming Degree}
\label{subsec:node_attributes}

Three node attributes are of central importance: the fitness $f_i$ of the optimum, its
self-loop weight $w_{ii}$, and its weighted incoming degree
$k_i^{\mathrm{in}} = \sum_{j \neq i} w_{ji}$. These three quantities are not
independent: they are jointly determined by basin geometry in ways that have direct
implications for innovation dynamics.

\subsubsection{Fitness and self-loop weight}

\begin{proposition}[Fitness--Self-Loop Correlation]
\label{prop:fitness_selfloop}
In fBm landscapes with persistence $\rho < 1$, fitness and self-loop weight are
positively correlated:
\begin{equation}
    \mathrm{Corr}(f_i,\, w_{ii}) > 0
\end{equation}
\end{proposition}

High-fitness optima in fBm landscapes sit atop broad, smooth hills where the
low-frequency octaves dominate. A random perturbation of radius $r$ from such a peak
is likely to land on the same slope and be returned to the same optimum by hill
climbing, yielding high $w_{ii}$. Low-fitness optima, by contrast, arise from
constructive interference of high-frequency octaves, producing narrow basins from
which perturbations easily escape. This correlation is the LON-theoretic foundation of
path dependence: high-performing technological approaches are inherently more stable,
and an agent that has found one will tend to return to it repeatedly. The reinforcement
arises endogenously from basin geometry rather than being imposed as a modeling
assumption.

\subsubsection{Fitness and incoming degree}

The weighted incoming degree $k_i^{\mathrm{in}} = \sum_{j \neq i} w_{ji}$ measures the
total probability flux arriving at $v_i$ from all other nodes in a single perturbation
step.

\begin{proposition}[Fitness--Incoming Degree Correlation]
\label{prop:fitness_indegree}
In fBm landscapes, fitness and weighted incoming degree are positively correlated:
\begin{equation}
    \mathrm{Corr}(f_i,\, k_i^{\mathrm{in}}) > 0
\end{equation}
\end{proposition}

Larger basins present a larger cross-section to perturbations from neighboring optima
and therefore accumulate higher incoming weight from more sources. Since basin size is
positively correlated with fitness by the same geometric argument as in
Proposition~\ref{prop:fitness_selfloop}, fitness and incoming degree inherit the same
positive correlation. High-fitness nodes are therefore simultaneously hard to escape
(high $w_{ii}$) and easy to fall into (high $k_i^{\mathrm{in}}$), so that the
stationary distribution of a random walk on the LON is concentrated on high-fitness
nodes.

\begin{figure}[H]
    \centering
    \includegraphics[width=\textwidth]{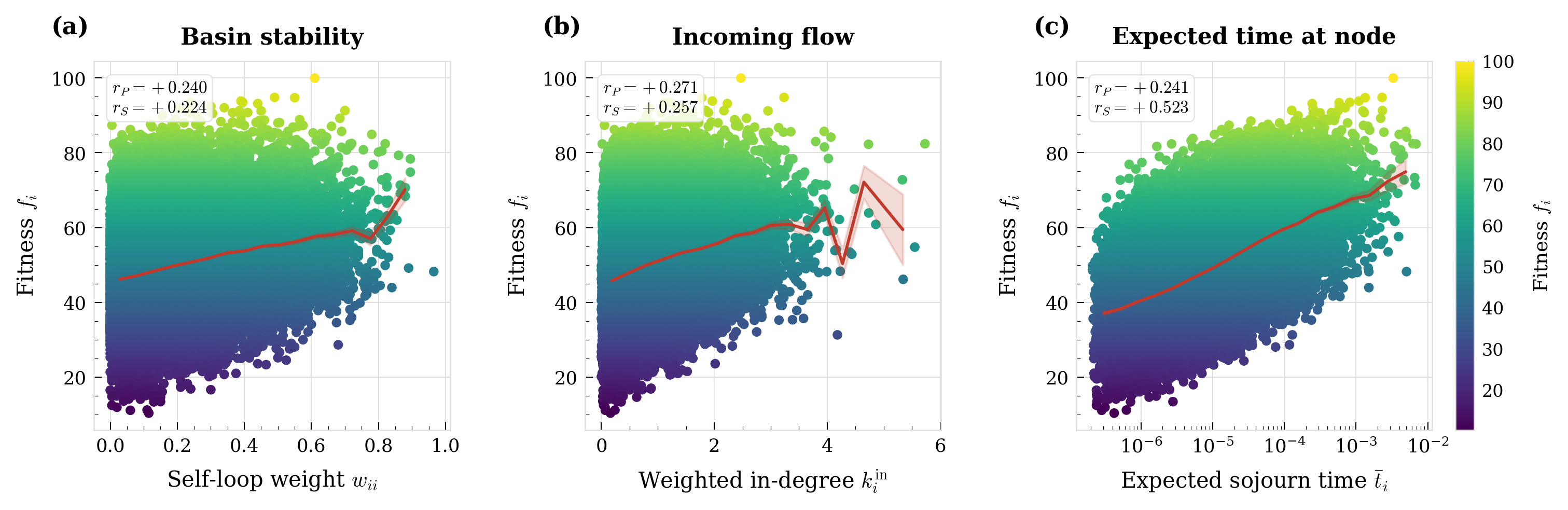}
    \caption{Node-level attribute relationships in the baseline LON
    ($\rho = 0.8$, $n = 7$, $\omega L = 600$). Left: fitness
    $f_i$ versus self-loop weight $w_{ii}$. Centre: fitness versus weighted incoming
    degree $k_i^{\mathrm{in}} = \sum_{j \neq i} w_{ji}$. Right: fitness versus
    expected sojourn time $\bar{t}_i = \pi_i / (1 - w_{ii})$ (log scale), where
    $\pi_i$ is the stationary probability of the random walk. Points are colored by
    fitness. Red lines show binned means ($\pm 1$ SE); Pearson
    ($r_P$) and Spearman ($r_S$) correlations are reported in each panel.}
    \label{fig:lon_node_attributes}
\end{figure}

\Cref{fig:lon_node_attributes} illustrates these relationships for the baseline LON
used throughout \Cref{sec:capabilities} ($\rho = 0.8$, $n = 7$, $\omega L = 600$).
The left panel shows fitness against self-loop weight ($r_S = 0.224$): the positive
trend is present but noisy. The centre panel shows fitness against weighted incoming
degree ($r_S = 0.257$): the trend is steeper and more structured. The right panel
shows fitness against expected sojourn time
$\bar{t}_i = \pi_i / (1 - w_{ii})$ on a logarithmic axis ($r_S = 0.523$): sojourn
times span roughly four orders of magnitude, from $\sim 10^{-6}$ for the most
transient low-fitness nodes to $\sim 10^{-2}$ for the dominant community attractors,
with an approximately log-linear relationship across this range.

\subsubsection{The joint structure and its consequences}

These relationships form a coherent joint structure. High-fitness nodes are simultaneously
large-basin, high self-loop, and high in-degree. They function as the stable attractors of
the network: easy to discover, hard to leave, and frequently revisited. Low-fitness nodes,
by contrast, sit in narrow basins with low self-loop weights, making them transient
waypoints from which agents are quickly displaced. The resulting heterogeneity in
residence times and visit frequencies is a structural feature of the LON that, as shown
in \Cref{sec:capabilities}, directly shapes the Zipf-law exponent and the heavy-tailed
character of inter-event times.

\subsection{Community Structure and Technological Paradigms}
\label{subsec:communities}

The node-level properties examined in \Cref{subsec:node_attributes} characterize
individual optima in isolation. A complementary question concerns the \emph{meso-scale}
structure of the LON: how do local optima organize into groups, and what do these groups
represent in innovation terms?

We identify communities using the \emph{Walktrap algorithm}
\citep{pons2005computing}, which groups nodes into clusters within which short random
walks tend to remain. The algorithm defines a distance between nodes based on
transition probabilities over short walks, then applies hierarchical agglomeration to
partition the network. Walktrap is a natural choice here because its underlying
dynamics are structurally identical to the walker model of \Cref{subsec:walker}, and
because it is sensitive to edge weights, so that the transition probabilities $w_{ij}$
directly inform community boundaries.

\begin{definition}[Technological Paradigm]
\label{def:paradigm}
A \emph{technological paradigm} in the LON model is operationalized as a
community $C_k \subset V$ identified
by Walktrap community detection: a set of local optima that are mutually accessible
through incremental basin transitions, but from which escape to other communities
requires either a rare large perturbation or an exogenous innovation event
($\varepsilon$-teleportation). Each community contains a high-fitness attractor node
that functions as the community's dominant stable configuration.
\end{definition}

Each community represents a coherent \emph{technological approach}: a family of related
designs among which an agent can move fluidly through basin transitions, with dynamics
governed by local optimization toward the community's high-fitness attractor. Crossing a
community boundary represents a qualitative reorientation that requires overcoming the
inter-community barriers encoded in the LON's edge structure. This provides a
network-grounded operationalization of the concept of technological paradigm
as used by \citet{dosi1982technological}: the community delineates the set of
configurations reachable through the paradigm's characteristic search heuristics, and
the community boundary marks where search must either make a large jump or wait for an
exogenous event to access a qualitatively different set of solutions.

\begin{figure}[H]
    \centering
    \includegraphics[width=\textwidth]{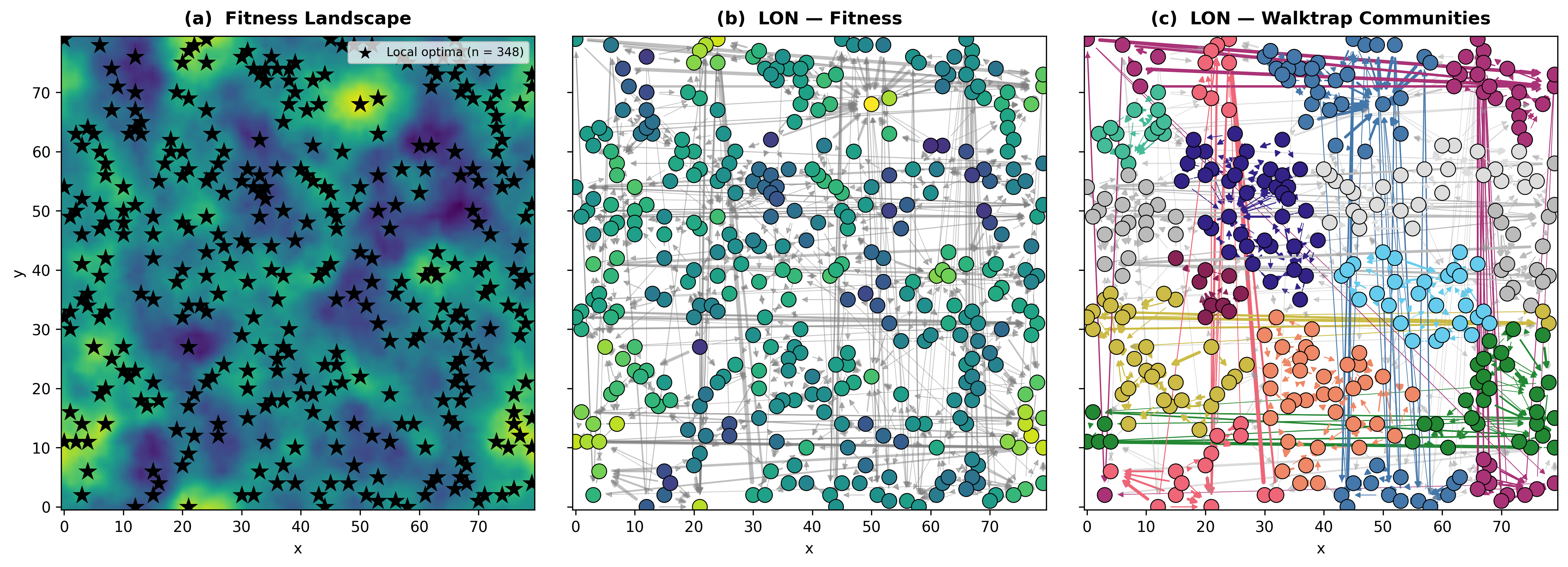}
    \caption{Community structure of a representative LON. (a) Fitness landscape
    with local optima marked as stars ($N = 622$). (b) LON with nodes colored
    by fitness $f_i$; edges shown in grey. (c) LON with nodes
    colored by Walktrap community assignment \citep{pons2005computing}. Each color
    represents a distinct community, interpreted as a technological paradigm: a set of
    locally stable approaches that are mutually accessible through incremental
    transitions. Edges crossing community boundaries are shown in the community
    color of their source node.}
    \label{fig:lon_communities}
\end{figure}

\Cref{fig:lon_communities} illustrates the community structure of a representative
LON. Communities are roughly spatially cohesive, reflecting that nearby optima tend to
be connected by short transitions, but long-range edges produced by toroidal wrapping
occasionally link spatially distant optima into the same community. Communities vary
substantially in size, with a few large communities dominating and many smaller
peripheral ones.

The community structure interacts with the node-level properties of
\Cref{subsec:node_attributes} in a systematic way. High-fitness nodes act as community
attractors, while low-fitness nodes near community boundaries are the most likely points
of inter-community transition. The resulting two-level vocabulary, basins for incremental
innovation and communities for technological paradigms, provides the interpretive
framework for \Cref{sec:capabilities}: residence within a basin corresponds to
exploitation of a specific configuration; movement across basins within a community
corresponds to incremental exploration within the current paradigm; and transitions
across community boundaries correspond to paradigm shifts.

 
\section{Model Capabilities: Replicating Innovation Regularities}
\label{sec:capabilities}
 
This section is the core analytical contribution of the paper. We show how the
LON-innovation model generates the four key empirical regularities of the
discovery-process tradition, identify the LON structural features responsible for
each, and demonstrate how the exponents respond to changes in model parameters.
 
\subsection{Baseline Configuration and a Representative Walk}
\label{subsec:baseline}
 
The baseline analysis uses the following parameter configuration: grid size
$L = 1{,}000$, base frequency $\omega = 0.6$ (so that $\omega L = 600$),
persistence $\rho = 0.8$, $n = 7$ octaves, 4-connected
neighborhood, perturbation radius $r = 10$, and $m = 200$ perturbation samples
per node. The exogenous innovation rate is set to $\varepsilon = 0$, so that all
exploration is driven entirely by the basin-hopping dynamics of the LON, without
any random teleportation.
 
\begin{figure}[H]
    \centering
    \includegraphics[width=\textwidth]{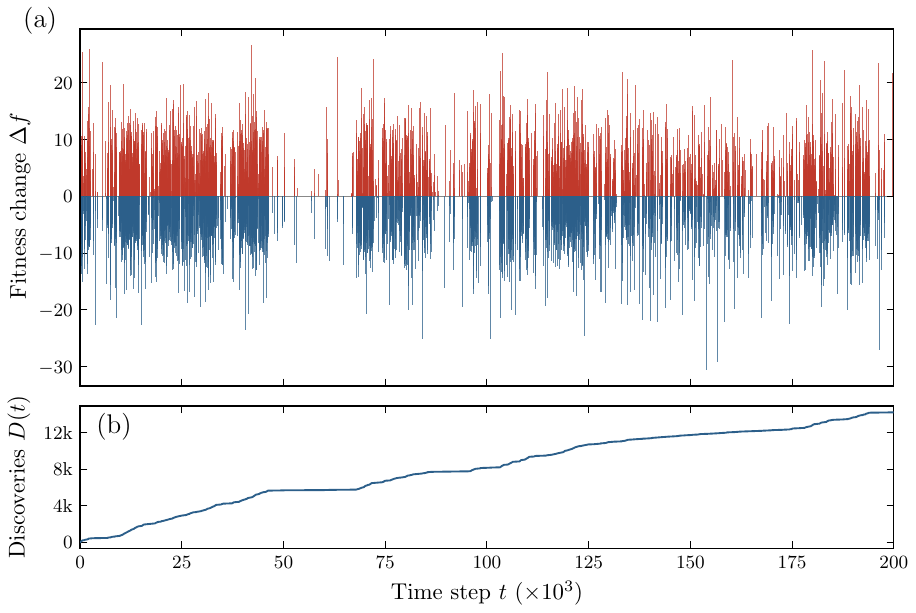}
    \caption{\textbf{Punctuated discovery dynamics.}
    A single walk on the baseline LON ($|V| = 47{,}272$ nodes).
    \textbf{(a)}~Fitness change $\Delta f = f_{\text{new}} - f_{\text{prev}}$
    at each discovery event.  Red (blue) spikes indicate fitness improvements
    (regressions); events cluster into bursts separated by quiescent intervals.
    \textbf{(b)}~Cumulative distinct optima $D(t)$. Plateaus correspond to trapping episodes; risers correspond to burst phases.}
    \label{fig:burstiness}
\end{figure}

Before examining ensemble statistics, it is instructive to look at what a single
walk on the LON actually produces. \Cref{fig:burstiness} shows a trajectory of
$T = 200{,}000$ steps on a LON with $|V| = 47{,}272$ nodes, constructed from the
baseline landscape parameters. Panel (a) records the fitness change
$\Delta f = f_{\text{new}} - f_{\text{prev}}$ at each discovery event, that is,
each time the walker visits a previously unvisited optimum, where $f_{\text{prev}}$
is the fitness of the last novel optimum discovered. Two features are immediately
apparent. Discovery events are not spread uniformly over time: they cluster into
dense bursts, separated by long quiescent intervals during which the walker revisits
already-known optima without encountering any new ones. And the fitness changes
within each burst are a mix of improvements (red spikes) and regressions (blue
spikes), reflecting the fact that the walker does not follow a monotonically improving
path through the LON but instead explores heterogeneous neighborhoods in which some
new optima have higher fitness than the previous discovery and others have lower.
 
Panel (b) shows the cumulative number of distinct optima discovered, $D(t)$. The
trajectory has a characteristic staircase shape: steep risers correspond to the burst
episodes visible in panel (a), during which the walker enters a new community or an
unexplored region of its current community and discovers many new optima in rapid
succession; plateaus correspond to the quiescent intervals during which the walker is
trapped at high self-loop nodes and the novelty count is flat. The plateaus grow longer as the walk progresses, because the walker
progressively exhausts the easily reachable optima in its vicinity and must traverse
increasingly rare inter-community transitions to find new ones.
 
This punctuated pattern is the single-trajectory signature of the statistical
regularities analyzed in the remainder of this section. The staircase shape of
$D(t)$ is what produces sublinear Heaps' law growth when averaged over many
replications. The clustering of discoveries into bursts is what produces heavy-tailed
inter-event time distributions. And the heterogeneous burst sizes across different
communities are what produce Zipf-law concentration in visit frequencies and
Taylor-law anomalous fluctuation scaling.
 
\begin{figure}[H]
    \centering
    \includegraphics[width=\textwidth]{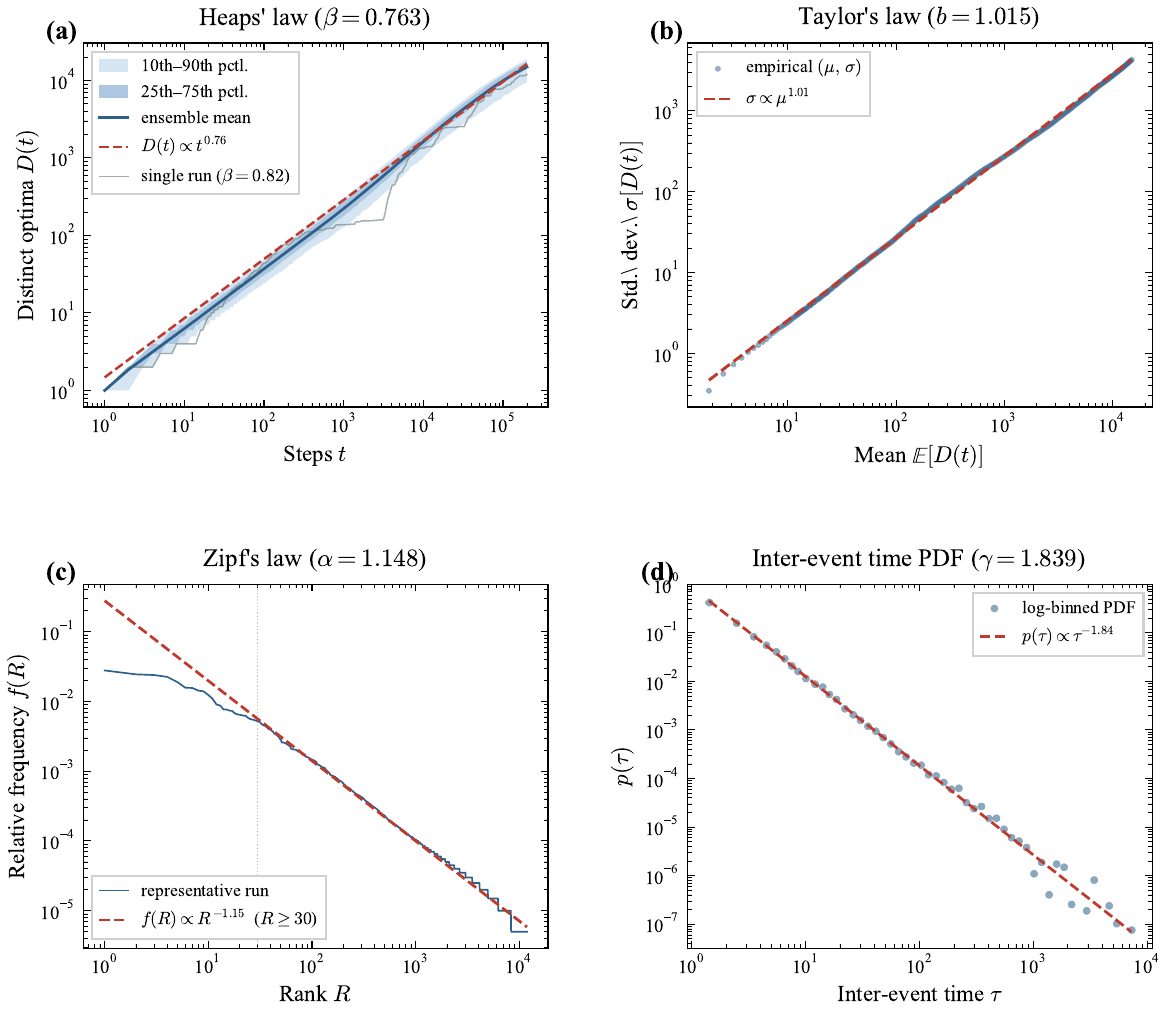}
    \caption{Baseline ensemble results ($\rho = 0.8$, $n = 7$ octaves,
    $\varepsilon = 0$; $20$ LONs $\times$ $50$ walks $= 1{,}000$ replications,
    $T = 200{,}000$).
    \textbf{(a)} Heaps' law: distinct optima $D(t)$ versus step $t$, with
    ensemble mean, percentile bands, and fitted exponent $\beta = 0.763$.
    \textbf{(b)} Taylor's law: $\sigma[D(t)]$ versus $\mathbb{E}[D(t)]$, with
    fitted exponent $b = 1.015$.
    \textbf{(c)} Zipf's law: rank-frequency distribution (representative run),
    with tail fit $\alpha = 1.148$ for ranks $R \geq 30$.
    \textbf{(d)} Inter-event time PDF (representative run), with fitted exponent
    $\gamma_{\mathrm{IET}} = 1.839$.}
    \label{fig:baseline}
\end{figure}

To characterize these regularities quantitatively, we adopt a two-level ensemble
design. We generate $20$ independent landscapes (and their corresponding LONs)
from different random seeds, and for each LON we run $50$ independent walker
replications of length $T = 200{,}000$ steps, yielding a total of $1{,}000$
replications. This design ensures that the reported statistics reflect variation
across both the structural properties of different LON realizations and the
stochastic trajectories of different walkers on the same LON.
\Cref{fig:baseline} summarizes the ensemble results.
 
\subsection{Heaps' Law: Sublinear Novelty Growth}
\label{subsec:heaps}
 
\Cref{fig:baseline}a shows the ensemble-averaged novelty count $D(t)$, the number
of distinct optima visited by step $t$, on log-log axes. The ensemble mean follows a
clear power-law regime $D(t) \sim t^{\beta}$ with fitted exponent $\beta = 0.763$,
and a representative single run ($\beta = 0.82$) is shown for comparison. The inner
shaded band covers the 25th--75th percentiles; the outer band spans the 10th--90th
percentiles across all $1{,}000$ replications.
 
The sublinear exponent $\beta < 1$ arises from the interplay between the LON's
self-loop structure and its community organization. At early times, the walker is
confined to its initial community, where it encounters new optima at a rate governed
by the local out-degree and the self-loop weights of the nodes it visits. Because
high-fitness nodes have high $w_{ii}$ (Proposition~\ref{prop:fitness_selfloop}), the
walker spends many steps returning to already-visited configurations before escaping to
a neighboring basin. This produces a discovery rate that is slower than linear from the
outset. As the walk progresses and the walker exhausts the optima in its current
community, further novelty requires crossing a community boundary, which is rare due to
the inter-community bottleneck structure. In the baseline configuration, where
$\varepsilon = 0$, there is no teleportation mechanism to assist escape: the walker
must reach a new community entirely through the basin-hopping dynamics. The decreasing
marginal rate of novelty production is therefore a direct consequence of the
hierarchical trapping structure of the LON: basins within communities, and communities
within the network as a whole, create nested barriers to exploration.
 
The fitted exponent $\beta = 0.763$ falls within the range
commonly reported in empirical innovation data
\citep{Tria2014CorrelatedNovelties,TriaLoretoServedio2018ZipfHeapsTaylor}. The
single-run exponent ($\beta = 0.82$) is somewhat higher, reflecting the greater
stochasticity of individual trajectories. The gap between single-run and ensemble
exponents is itself informative: it indicates that averaging over multiple LON
realizations and their heterogeneous community structures produces a more regular,
and somewhat slower, novelty growth curve than any individual realization.

\subsection{Taylor's Law: Fluctuation Scaling}
\label{subsec:taylor}
 
\Cref{fig:baseline}b shows the Taylor-law relationship between the mean $\mu$ and
standard deviation $\sigma$ of the cumulative novelty count $D(t)$ across the
ensemble, evaluated at logarithmically spaced time points. The empirical relationship
$\sigma \sim \mu^{b}$ is fitted with exponent $b = 1.015$.
 
The exponent $b \approx 1$ indicates that fluctuations in the novelty count scale
almost linearly with the mean, placing the model firmly in the stronger-than-Poisson,
correlated regime. This is a direct consequence of the two-level heterogeneity built
into the ensemble design. At the first level, different LON realizations have different
community structures, basin size distributions, and self-loop weight profiles, so that
walkers on different LONs face structurally different exploration landscapes. At the
second level, walkers on the same LON starting from different initial conditions can
become trapped in different communities for extended periods. A walker that arrives
early in a large, high-connectivity community discovers many new optima quickly, while
a walker trapped in a small community with high self-loop weights at its attractor
node discovers new optima slowly. This community-level heterogeneity in discovery
rates amplifies the cross-replication variance of $D(t)$ beyond what a memoryless
process would produce.
 
The Taylor exponent $b \approx 1$ is consistent with the values reported in empirical
studies of innovation records
\citep{TriaLoretoServedio2018ZipfHeapsTaylor,TriaCrimaldiAlettiServedio2020TaylorInnovation}
and with the theoretical prediction that correlated, bursty discovery processes produce
Taylor exponents near or above unity.

\subsection{Zipf's Law: Heavy-Tailed Reuse}
\label{subsec:zipf}
 
\Cref{fig:baseline}c shows the rank-frequency distribution of visited optima for a
representative run: the relative visit frequency $f(R)$ of the optimum of rank $R$
(sorted by decreasing frequency) on log-log axes. The tail of the distribution (ranks
$R \geq 30$) follows a power law $f(R) \sim R^{-\alpha}$ with fitted exponent
$\alpha = 1.148$.
 
The Zipf exponent close to unity reflects the concentration of visits on a small
number of high-fitness, high self-loop nodes. As established in
\Cref{subsec:node_attributes}, high-fitness nodes simultaneously attract large incoming
flow ($k_i^{\mathrm{in}}$ high) and retain walkers for long sojourns ($w_{ii}$ high).
The stationary distribution of the random walk is therefore heavily skewed: a few
community-attractor nodes accumulate a disproportionate share of the total visit mass,
while the many low-fitness transient nodes in the network periphery are visited rarely
and briefly.
 
The characteristic shape of the rank-frequency curve, with a flattened head at low
ranks and a steeper power-law tail, is a signature of the LON's joint structure of
fitness, self-loop weight, and incoming degree. The top-ranked optima (roughly the
top 20--30) correspond to the dominant attractors of the largest communities; their
visit frequencies are compressed relative to one another because even without
teleportation, the basin-hopping dynamics eventually carry the walker across community
boundaries, redistributing some visit mass. Beyond the head, the tail follows the
power law: the many small-basin, low-fitness optima that form the periphery of each
community are visited with frequencies that decay as a regular function of rank.

\subsection{Inter-Event Time Distribution: Bursty Discovery Dynamics}
\label{subsec:iet}
 
\Cref{fig:baseline}d shows the probability density function of inter-event times
$\tau$, where $\tau$ is the number of steps between consecutive novelty events (first
visits to previously unvisited optima), for a representative run. The log-binned
empirical density follows a power law
$p(\tau) \sim \tau^{-\gamma_{\mathrm{IET}}}$ with fitted exponent
$\gamma_{\mathrm{IET}} = 1.839$ over more than three orders of magnitude in $\tau$.
 
The heavy-tailed inter-event time distribution is the temporal fingerprint of the
LON's community and trapping structure. Within a community, novelty events can occur
in rapid succession as the walker transitions between neighboring basins, producing
short inter-event times. Once the walker has visited most optima in its current
community, however, further novelty requires a rare inter-community transition
through the basin-hopping dynamics alone (since $\varepsilon = 0$). The walker then
enters a long quiescent phase during which it revisits already-known optima, producing
a long inter-event time. The distribution of these quiescent durations is governed by
the distribution of self-loop sojourn times and community escape times, both of which
are heavy tailed due to the heterogeneous $w_{ii}$ distribution across nodes and the
heterogeneous community sizes across the LON.
 
The resulting picture is one of bursty discovery: episodes of rapid novelty production
when the walker enters a new community (or a previously unexplored region of its
current community), separated by long waiting times during which the walker is trapped
at high-fitness attractors. This is the temporal analog of the spatial heterogeneity
captured by Zipf's law: just as visit frequencies are concentrated on a few dominant
optima, novelty events are concentrated in temporal bursts rather than spread uniformly
over time.
 
The exponent $\gamma_{\mathrm{IET}} = 1.839$ falls in the range commonly observed in empirical studies.

\subsection{Joint Consistency of the Four Regularities}
\label{subsec:joint}
 
A key feature of the results presented above is that all four regularities emerge
simultaneously from the same ensemble of walks on the same set of LONs, without any
parameter tuning specific to individual regularities. The exponents are jointly
determined by the structural properties of the LON: the distribution of self-loop
weights, the community size distribution, and the fitness-degree correlation.
 
The joint consistency reflects the fact that the LON encodes a single, coherent
trapping-and-escape landscape, and the four regularities are simply different
projections of the same underlying dynamics. Heaps' law describes the cumulative
count of distinct states; Zipf's law describes the equilibrium distribution of visits
across states; Taylor's law describes the cross-replication variability of cumulative
counts; and the inter-event time distribution describes the temporal fine structure of
novelty production. All four are governed by the same competition between
within-community exploitation (driven by self-loop weights and community attractors)
and cross-community exploration (driven by low-fitness boundary nodes and the
basin-hopping dynamics).

\subsection{Parameter Variations}
\label{subsec:param_variations}
 
The baseline configuration is not the only one that produces the four regularities.
To illustrate how the exponents respond to changes in model parameters, we consider
three variations that each modify a single aspect of the baseline while holding
everything else fixed. \Cref{tab:param_variations} summarizes the four
configurations and their fitted exponents;
\Cref{fig:param_octaves,fig:param_epsilon,fig:param_persistence} show the
corresponding ensemble panels.
 
\begin{table}[H]
\centering
\caption{Parameter configurations and fitted exponents. All configurations use
$L = 1{,}000$, 4-connected neighborhood, $r = 10$, $m = 200$,
$T = 200{,}000$, $20$ LONs $\times$ $50$ walks.}
\label{tab:param_variations}
\begin{tabular}{@{}lccccccc@{}}
\toprule
\textbf{Configuration} & $\rho$ & $n$ & $\varepsilon$ &
    $\beta$ & $b$ & $\alpha$ & $\gamma_{\mathrm{IET}}$ \\
\midrule
Baseline                     & 0.8 & 7 & $0$         & 0.763 & 1.015 & 1.148 & 1.839 \\
Fewer octaves                & 0.8 & 6 & $0$         & 0.650 & 1.113 & 1.223 & 1.492 \\
Higher persistence           & 0.9 & 7 & $0$         & 0.732 & 1.091 & 0.950 & 1.922 \\
Exogenous innovation         & 0.8 & 7 & $10^{-5}$   & 0.777 & 0.990 & 1.066 & 1.867 \\
\bottomrule
\end{tabular}
\end{table}

\subsubsection{Reducing landscape complexity: fewer octaves}
 
\begin{figure}[H]
    \centering
    \includegraphics[width=\textwidth]{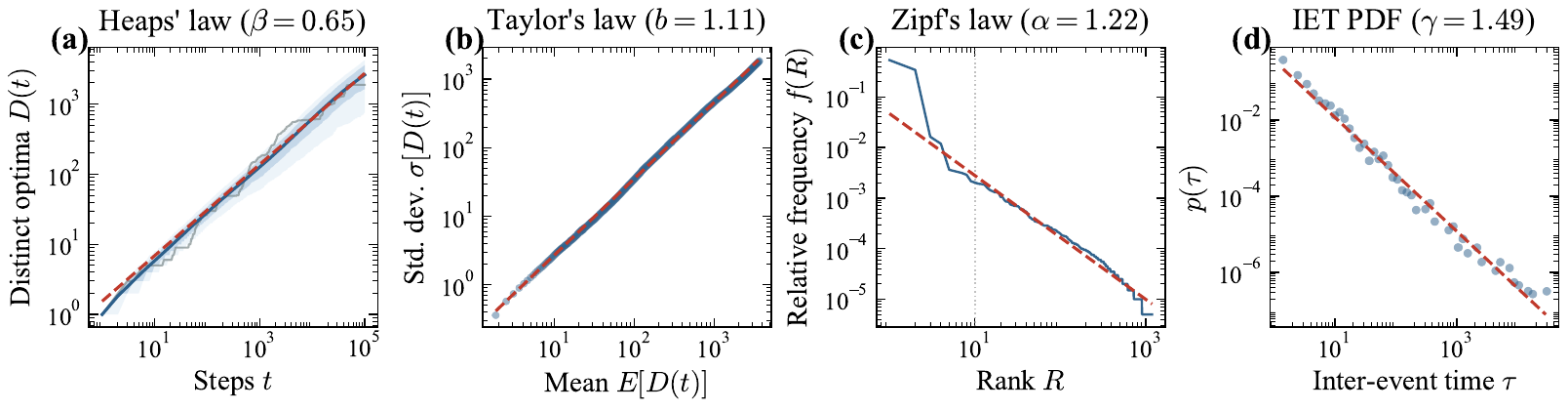}
    \caption{Ensemble results with reduced landscape complexity ($n = 6$ octaves,
    $\varepsilon = 0$; all other parameters as in \Cref{fig:baseline}).
    Fitted exponents: $\beta = 0.650$, $b = 1.113$, $\alpha = 1.223$,
    $\gamma_{\mathrm{IET}} = 1.492$.}
    \label{fig:param_octaves}
\end{figure}

\Cref{fig:param_octaves} shows the ensemble results when the number of octaves is
reduced from $n = 7$ to $n = 6$, with all other parameters held at their baseline
values. Removing the highest-frequency layer produces a smoother landscape with fewer
local optima and larger basins, yielding a LON with fewer nodes, broader basins, and
higher average self-loop weights.
 
The consequences are systematic: the Heaps exponent drops to $\beta = 0.650$ (slower
novelty growth due to stronger trapping in broad basins); the Zipf exponent rises to
$\alpha = 1.223$ (more concentrated visits on the fewer dominant attractors); the
Taylor exponent rises to $b = 1.113$ (greater cross-replication variability because
the starting community matters more when communities are few); and the inter-event
time exponent drops to $\gamma_{\mathrm{IET}} = 1.492$, indicating a substantially
heavier tail. This last change is the most striking: with fewer and broader basins the
self-loop sojourn times at dominant attractors are longer and inter-community
transitions are rarer, making the landscape less navigable and quiescent intervals
longer.

\subsubsection{Increasing fine-scale ruggedness: higher persistence}
 
\begin{figure}[H]
    \centering
    \includegraphics[width=\textwidth]{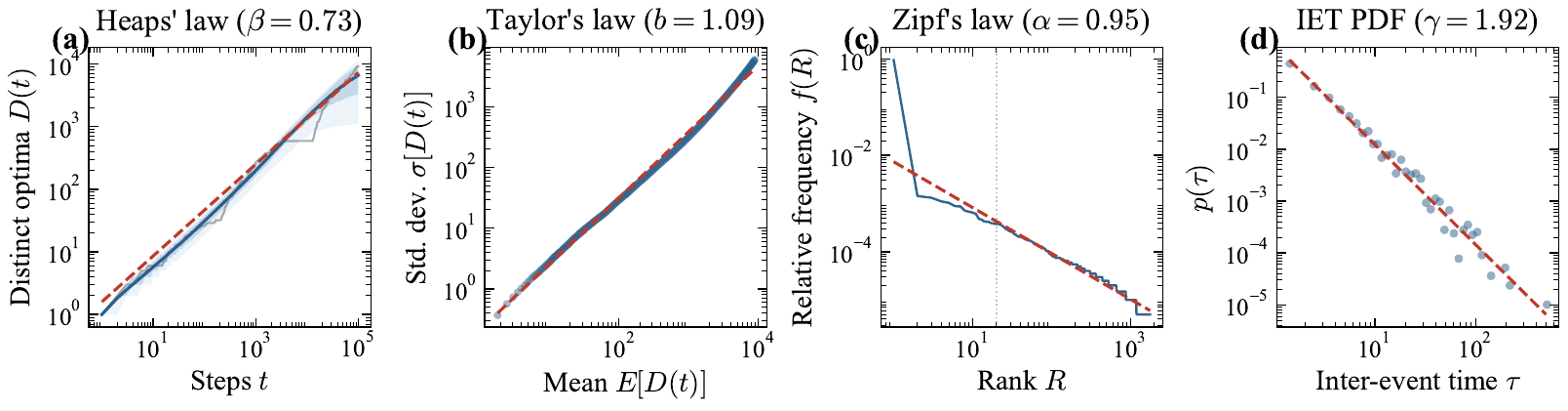}
    \caption{Ensemble results with higher persistence ($\rho = 0.9$,
    $\varepsilon = 0$; all other parameters as in \Cref{fig:baseline}).
    Fitted exponents: $\beta = 0.732$, $b = 1.091$, $\alpha = 0.950$,
    $\gamma_{\mathrm{IET}} = 1.922$.}
    \label{fig:param_persistence}
\end{figure}

\Cref{fig:param_persistence} shows the ensemble results when persistence is increased
to $\rho = 0.9$, with all other parameters at baseline. Higher $\rho$ up-weights
the high-frequency octaves, producing a more rugged landscape with more local optima,
narrower basins, and a denser, more heterogeneous edge structure.
 
The effects are in several respects the opposite of the fewer-octaves case. The Heaps
exponent decreases modestly to $\beta = 0.732$, but here the slowdown arises not from
fewer optima but from denser local cycling among many small basins without substantial
progress toward unexplored communities. The Zipf exponent drops below unity to
$\alpha = 0.950$, reflecting the proliferation of small basins that dilutes the
dominance of top-ranked attractors. The Taylor exponent rises to $b = 1.091$ because
the larger number of heterogeneous communities makes each walker's early trajectory
more consequential. The inter-event time exponent increases to
$\gamma_{\mathrm{IET}} = 1.922$, indicating a slightly lighter tail: denser local
structure provides more inter-basin transitions within each community, shortening
typical waiting times even as the overall discovery rate is slower.

The contrast between the two variations is instructive. Both slow novelty growth,
but through opposite structural mechanisms: removing fine-scale detail creates deeper
traps, while adding it creates denser local cycling. Their signatures in the remaining
exponents diverge accordingly: fewer octaves concentrates visits ($\alpha$ up) and
produces heavier inter-event tails ($\gamma_{\mathrm{IET}}$ down), while higher
persistence disperses visits ($\alpha$ down) and produces lighter tails
($\gamma_{\mathrm{IET}}$ up). This confirms that the four exponents are jointly
constrained by LON structure and that different routes to slower novelty growth leave
distinct fingerprints in the other three regularities.

\subsubsection{Introducing exogenous innovation}
 
\begin{figure}[H]
    \centering
    \includegraphics[width=\textwidth]{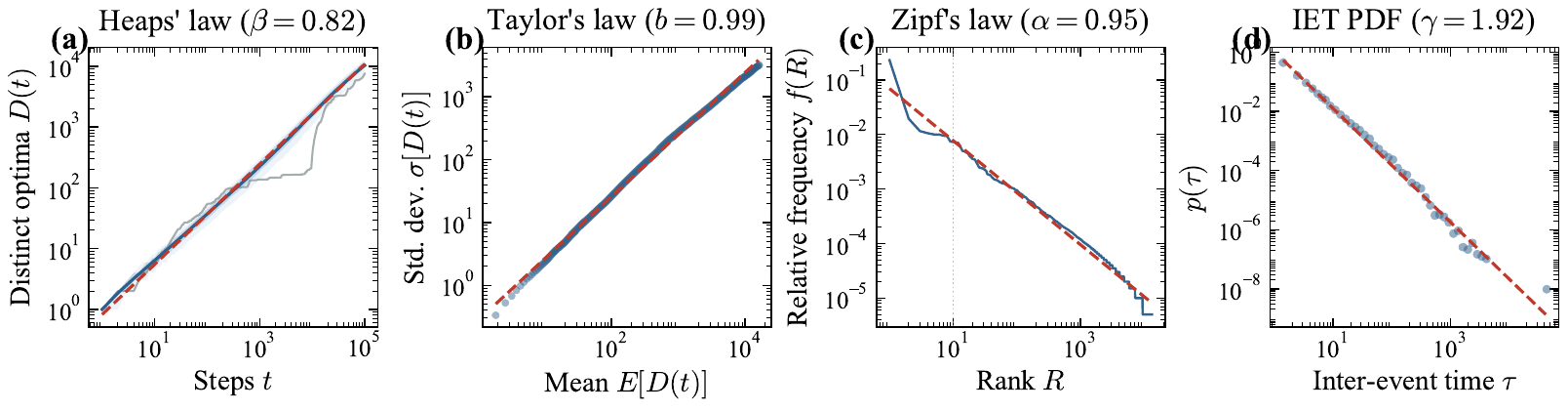}
    \caption{Ensemble results with exogenous innovation ($\varepsilon = 10^{-5}$,
    $n = 7$ octaves; all other parameters as in \Cref{fig:baseline}).
    Fitted exponents: $\beta = 0.777$, $b = 0.990$, $\alpha = 1.066$,
    $\gamma_{\mathrm{IET}} = 1.867$.}
    \label{fig:param_epsilon}
\end{figure}

\Cref{fig:param_epsilon} shows the ensemble results when a small exogenous innovation
rate $\varepsilon = 10^{-5}$ is introduced, with all other parameters at baseline.
At each step the walker now has a probability $\varepsilon$ of teleporting to a
uniformly random node, regardless of the current position or local edge structure.
 
The effects are modest but interpretable. Novelty growth is marginally faster
($\beta = 0.777$) because occasional random jumps bypass inter-community bottlenecks.
Visit concentration decreases ($\alpha = 1.066$) as teleportation redistributes mass
more evenly. Cross-replication variability decreases slightly ($b = 0.990$) because all
walkers eventually sample a broad cross-section of the LON. The inter-event time
exponent increases marginally ($\gamma_{\mathrm{IET}} = 1.867$) as teleportation
truncates the longest quiescent intervals. At $\varepsilon = 10^{-5}$ and
$T = 200{,}000$ steps the expected number of teleportation events per walk is only two,
yet this is sufficient to prevent the most extreme forms of lock-in. Exogenous
innovation thus acts as a weak regularizer: it smooths the most extreme consequences
of trapping without altering the qualitative character of the innovation record.

 
\section{Discussion}
\label{sec:discussion}
 
The LON-innovation model developed in this paper demonstrates that a single structural
object, the Local Optima Network, can generate the four main empirical regularities of the
discovery-process tradition while retaining the basin-level, performance-ordered
representation characteristic of the adaptive-search tradition. The parameter variations
of \Cref{subsec:param_variations} show, moreover, that the four exponents respond to
changes in landscape ruggedness ($\rho$), landscape complexity ($n$), and exogenous
innovation ($\varepsilon$) in interpretable, jointly constrained ways. This section
discusses the scope conditions of the current framework, its limitations, and the most
promising directions for extension.
 
\subsection{Scope and Limitations of the Baseline Model}
 
The baseline model makes several simplifying choices that should be made explicit. The
landscape is two-dimensional, which preserves all qualitative features of interest (multiple
optima, basins, communities, ruggedness variation) but limits the combinatorial richness
of the configuration space relative to the high-dimensional binary strings of NK models.
The walker dynamics are Markovian: the transition probability at each step depends only on
the current node, not on the history of the walk. This rules out reinforcement effects, memory,
and learning within the current formulation. The baseline operates with $\varepsilon = 0$,
and even when exogenous innovation is introduced, the $\varepsilon$-teleportation mechanism
imposes a structureless form of it: jumps are uniformly random and independent of the
walker's current fitness, the fitness of potential targets, or the behavior of other agents.
Each of these simplifications points toward a natural extension.
 
\subsection{Alternative Walker Dynamics}
 
The baseline model deliberately adopts the simplest possible dynamics: a memoryless
random walker whose transition probabilities are given directly by the basin-hopping
weights $w_{ij}$. This isolates the contribution of LON topology to the observed
regularities without introducing additional behavioral parameters.

Several richer dynamics are worth exploring. A \emph{temperature-modulated walker}
would bias transitions toward higher-fitness neighbors, with a temperature parameter
interpolating between the baseline uniform dynamics and greedy climbing on the LON.
This would be relevant for settings in which innovating agents preferentially pursue
improvements, potentially producing tighter visit concentration and longer trapping
episodes. An \emph{edge-reinforced random walk}
\citep{Iacopini2018NetworkDynamicsInnovation} would update transition probabilities as
a function of past traversals, generating endogenous preferential reuse and a form of
learning or habit formation. The results of \Cref{sec:capabilities} show, however,
that the baseline walker already generates all four regularities within empirically
observed ranges, validating the choice of starting with a minimal behavioral model.
 
\subsection{From Isolated Walkers to Interacting Agents}
 
The baseline model considers a single walker in isolation. In real innovation systems,
firms, inventors, and research teams observe one another, imitate successful strategies,
and compete for the same performance peaks. A natural extension is an agent-based model
(ABM) in which multiple walkers navigate the same LON simultaneously, with the
$\varepsilon$-teleportation mechanism replaced by a fitness-directed imitation rule:
an agent observes the current positions and fitness levels of other agents and, with
some probability proportional to the fitness differential, jumps toward a
better-performing peer. This endogenizes the direction of long-range jumps, introduces
a social dimension to exploration, and generates a natural coupling between the
discovery processes of different agents.

The ABM extension also opens the door to studying the distinction between local and
system-level novelty. With multiple walkers, a configuration can be new to one agent
but already known to others, and the rate at which system-level novelty grows will
depend on the degree of overlap in exploration trajectories. Imitation tends to increase
this overlap, creating a trade-off between individual efficiency and collective diversity
of exploration that is central to the organizational search literature
\citep{march1991exploration,BaumannEtAl2019}.
 
\subsection{Multi-Landscape Dynamics and Radical Innovation}
 
The baseline model operates on a single landscape, with communities serving as the
operational definition of distinct technological paradigms. However, the history of
technology also features episodes of radical discontinuity in which an entirely new
performance landscape emerges, rendering the old one obsolete: the transition from
vacuum tubes to semiconductors, from film to digital photography, or from internal
combustion to electric propulsion. A natural extension is a multi-landscape model in
which the walker (or population of walkers) can transition from the current landscape
$\mathcal{L}^{(k)}$ to a newly generated landscape $\mathcal{L}^{(k+1)}$ with a
potentially higher performance ceiling. The re-initialization of agents at low-fitness
positions in the new landscape would capture the Schumpeterian logic of creative
destruction \citep{Schumpeter1939BusinessCycles,dosi1982technological}. A sequence of
landscapes with progressively higher ceilings would also allow the model to generate
long-run performance growth \citep{nelson1982evolutionary}, connecting the short-run
regularities studied here to the long-run trajectories studied in the technological
change literature.
 
\subsection{Empirical Strategies}
 
The model generates several testable predictions that could, in principle, be confronted with
data. The most direct empirical strategy is to estimate the four exponents ($\beta$, $\alpha$,
$b$, $\gamma_{\mathrm{IET}}$) from patent, publication, or product-innovation data and compare
them to the exponent ranges produced by the model under different parameterizations. The
parameter variations of \Cref{tab:param_variations} show that the model generates a
non-trivial joint structure among the four exponents: for instance, smoother landscapes
(fewer octaves) slow novelty growth while concentrating visits and producing heavier
inter-event tails, whereas more rugged landscapes (higher persistence) also slow novelty
growth but disperse visits and produce lighter tails. These distinct joint signatures
provide sharper tests than would be available from any single exponent alone.
A more structural approach would attempt to reconstruct the LON, or at least its community structure
and degree distribution, from observed sequences of technological choices. Citation networks
among patents \citep{Price1965CitationNetworks}, recombination networks among components
\citep{FlemingSorenson2001Recombination}, and co-occurrence networks among technological
codes \citep{AharonsonSchilling2016TechLandscape} all provide empirical network representations
of technological relatedness and knowledge flow that may serve as partial proxies for the
basin-transition structure that the LON formalizes.
 
\subsection{Alternative Landscape Generation Mechanisms}
 
The LON framework is not tied to the fBm landscape generator used here. A natural
alternative is the NK model \citep{kauffman1993origins,levinthal1997adaptation}, the
standard benchmark in the organizational search literature, which offers tunable
ruggedness through the epistasis parameter $K$ and fine-grained control over
interdependence structure \citep{Frenken2006Modularity,ethiraj2004modularity}. The LON
construction procedure of \Cref{subsec:lon} applies without modification to NK
landscapes, and the resulting NK-LONs have been extensively characterized in the
optimization literature \citep{ochoa2008study,verel2011local,ochoa2014lon}. Comparing
innovation-record statistics from fBm-LONs versus NK-LONs would establish the
robustness of the regularities documented in \Cref{sec:capabilities} to the choice of
generator. Beyond NK, other generators of interest include block models with modular
interaction structures, random field models with long-range correlations, and
empirically calibrated landscapes constructed from patent or product data.
 
\subsection{Higher-Dimensional Landscapes and Computational Scalability}
 
The two-dimensional configuration space used throughout this paper enables tractable
exhaustive enumeration of all local optima and their basins, but real technological design
spaces are vastly higher-dimensional. Extending the model to $d > 2$ dimensions is
conceptually straightforward, since the Clifford torus mapping and fBm construction
generalize naturally, but computationally demanding: exhaustive hill climbing from every grid
point scales as $L^d$, making full enumeration infeasible for large $d$. Sampling-based LON
construction methods, including the Markov-chain and snowball samplers discussed in
\Cref{subsec:lit_lons}, offer a path forward, but the relationship between sampled and
exhaustive LONs in high dimensions requires further study.

 
\section{Conclusion}
\label{sec:conclusion}
 
This paper has introduced a model of innovation dynamics grounded in Local Optima Networks.
The model constructs a LON from a toroidal fitness landscape generated via four-dimensional
Perlin noise, identifies communities in the LON as technological paradigms, and studies
stochastic walkers on the resulting network as a representation of innovating agents.
 
The central result is that walks on LONs simultaneously generate the four main empirical
regularities of the discovery-process tradition: sublinear novelty growth (Heaps' law),
heavy-tailed rank-frequency distributions (Zipf's law), anomalous fluctuation scaling
(Taylor's law), and power-law distributed inter-event times, with fitted exponents that
fall within the ranges commonly reported in empirical innovation data. All four emerge
from the same structural features of the LON: the heterogeneous distribution of self-loop
weights, the community organization of basins into distinct technological paradigms, and
the fitness-degree correlations that concentrate visits on a small number of
high-performance attractors. The model thereby provides a bridge between the
discovery-process and adaptive-search traditions, connecting the sequence-level
statistical regularities emphasized by the former to the basin-level,
performance-ordered landscape structure emphasized by the latter.
 
Parameter variations show that the exponents are jointly constrained by the LON's
structural properties and respond to changes in interpretable, structurally grounded ways.
Different routes to the same macroscopic outcome leave distinct signatures in the
remaining exponents, making the model's predictions falsifiable at the level of exponent
vectors, not just individual scaling laws.
 
The framework is parsimonious: the landscape is controlled by a small number of
parameters, the LON construction is governed by the perturbation radius, and the walker
dynamics require only an optional exogenous innovation rate, which can be set to zero
without losing any of the four regularities. Yet despite this parsimony, the model
admits natural extensions to multi-agent settings with fitness-directed imitation,
to multi-landscape settings with radical technological transitions, to alternative
walker dynamics such as temperature-biased or edge-reinforced walks, and to alternative
landscape generators including NK models.
 
The LON representation does not replace either the discovery-process or the adaptive-search
tradition. It provides a common formal object through which their distinct explanatory targets
can be connected, and it makes the structural assumptions underlying innovation dynamics
explicit, interpretable, and amenable to empirical testing.

\bibliography{references}

@article{Tria2014CorrelatedNovelties,
  author  = {Tria, Francesca and Loreto, Vittorio and Servedio, Vito D. P. and Strogatz, Steven H.},
  title   = {The dynamics of correlated novelties},
  journal = {Scientific Reports},
  volume  = {4},
  pages   = {5890},
  year    = {2014}
}

@article{TriaLoretoServedio2018ZipfHeapsTaylor,
  author  = {Tria, Francesca and Loreto, Vittorio and Servedio, Vito D. P.},
  title   = {Zipf's, Heaps' and Taylor's Laws are Determined by the Expansion into the Adjacent Possible},
  journal = {Entropy},
  volume  = {20},
  number  = {10},
  pages   = {752},
  year    = {2018}
}

@article{TriaCrimaldiAlettiServedio2020TaylorInnovation,
  author  = {Tria, Francesca and Crimaldi, Irene and Aletti, Giovanni and Servedio, Vito D. P.},
  title   = {Taylor's Law in Innovation Processes},
  journal = {Entropy},
  volume  = {22},
  number  = {5},
  pages={573},
  year    = {2020},
  doi     = {10.3390/e22050573}
}

@article{Iacopini2018NetworkDynamicsInnovation,
  author  = {Iacopini, I. and Milojevi{\'c}, S. and Latora, V.},
  title   = {Network dynamics of innovation processes},
  journal = {Physical Review Letters},
  volume  = {120},
  pages   = {048301},
  year    = {2018}
}

@article{marzo2022modeling,
  title={Modeling innovation in the cryptocurrency ecosystem},
  author={De Marzo, Giordano and Pandolfelli, Francesco and Servedio, Vito DP},
  journal={Scientific Reports},
  volume={12},
  number={1},
  pages={12942},
  year={2022},
  publisher={Nature Publishing Group UK London}
}

@book{Schumpeter1939BusinessCycles,
  author    = {Schumpeter, Joseph A.},
  title     = {Business Cycles},
  volume    = {1},
  publisher = {McGraw-Hill},
  year      = {1939}
}

@article{dosi1982technological,
  author  = {Dosi, Giovanni},
  title   = {Technological paradigms and technological trajectories: A suggested interpretation of the determinants and directions of technical change},
  journal = {Research Policy},
  year    = {1982},
  volume  = {11},
  number  = {3},
  pages   = {147--162}
}

@article{frenken2006technological,
  author  = {Frenken, Koen},
  title   = {Technological innovation and complexity theory},
  journal = {Economics of Innovation and New Technology},
  year    = {2006},
  volume  = {15},
  number  = {2},
  pages   = {137--155}
}

@book{kauffman1993origins,
  author    = {Kauffman, Stuart A.},
  title     = {The Origins of Order: Self-Organization and Selection in Evolution},
  publisher = {Oxford University Press},
  year      = {1993}
}

@article{levinthal1997adaptation,
  author  = {Levinthal, Daniel A.},
  title   = {Adaptation on rugged landscapes},
  journal = {Management Science},
  year    = {1997},
  volume  = {43},
  number  = {7},
  pages   = {934--950}
}

@article{march1991exploration,
  author  = {March, James G.},
  title   = {Exploration and exploitation in organizational learning},
  journal = {Organization Science},
  year    = {1991},
  volume  = {2},
  number  = {1},
  pages   = {71--87}
}

@book{nelson1982evolutionary,
  author    = {Nelson, Richard R. and Winter, Sidney G.},
  title     = {An Evolutionary Theory of Economic Change},
  publisher = {Harvard University Press},
  year      = {1982}
}

@inproceedings{ochoa2008study,
  author    = {Ochoa, Gabriela and Tomassini, Marco and V{\'e}rel, S{\'e}bastien and Darabos, Christian},
  title     = {A study of {NK} landscapes' basins and local optima networks},
  booktitle = {Proceedings of the 10th Annual Conference on Genetic and Evolutionary Computation (GECCO)},
  year      = {2008},
  pages     = {555--562}
}

@article{perlin1985image,
  author  = {Perlin, Ken},
  title   = {An image synthesizer},
  journal = {ACM SIGGRAPH Computer Graphics},
  year    = {1985},
  volume  = {19},
  number  = {3},
  pages   = {287--296}
}

@article{verel2011local,
  author  = {V{\'e}rel, S{\'e}bastien and Ochoa, Gabriela and Tomassini, Marco},
  title   = {Local optima networks of {NK} landscapes with neutrality},
  journal = {IEEE Transactions on Evolutionary Computation},
  year    = {2011},
  volume  = {15},
  number  = {6},
  pages   = {783--797}
}

@inproceedings{wright1932roles,
  author    = {Wright, Sewall},
  title     = {The roles of mutation, inbreeding, crossbreeding, and selection in evolution},
  booktitle = {Proceedings of the Sixth International Congress of Genetics},
  year      = {1932},
  volume    = {1},
  pages     = {356--366}
}

@article{khraisha2020complex,
  title   = {Complex economic problems and fitness landscapes: Assessment and methodological perspectives},
  author  = {Khraisha, T.},
  journal = {Structural Change and Economic Dynamics},
  volume  = {52},
  pages   = {390--407},
  year    = {2020},
  doi     = {10.1016/j.strueco.2019.01.002}
}

@article{ethiraj2004modularity,
  title   = {Modularity and innovation in complex systems},
  author  = {Ethiraj, Sendil K. and Levinthal, Daniel},
  journal = {Management Science},
  volume  = {50},
  pages   = {159--173},
  year    = {2004}
}

@misc{ochoa2014lon,
    author="Ochoa, Gabriela
    and Verel, S{\'e}bastien
    and Daolio, Fabio
    and Tomassini, Marco",
    editor="Richter, Hendrik
    and Engelbrecht, Andries",
    title="Local Optima Networks: A New Model of Combinatorial Fitness Landscapes",
    bookTitle="Recent Advances in the Theory and Application of Fitness Landscapes",
    year="2014",
    publisher="Springer Berlin Heidelberg",
    address="Berlin, Heidelberg",
    pages="233--262",
    }

@inproceedings{chicano2012autocorrelation,
  title     = {Local optima networks, landscape autocorrelation and heuristic search performance},
  author    = {Chicano, Francisco and Daolio, Fabio and Ochoa, Gabriela and Verel, S{\'e}bastien and Tomassini, Marco and Alba, Enrique},
  booktitle = {Parallel Problem Solving from Nature -- PPSN XII},
  series    = {Lecture Notes in Computer Science},
  volume    = {7492},
  pages     = {337--347},
  publisher = {Springer},
  year      = {2012}
}

@article{pons2005computing,
  author  = {Pons, Pascal and Latapy, Matthieu},
  title   = {Computing communities in large networks using random walks},
  journal = {Journal of Graph Algorithms and Applications},
  volume  = {10},
  number  = {2},
  pages   = {191--218},
  year    = {2006}
}

@inproceedings{CleghornOchoa2021PSOLON,
  author    = {Cleghorn, Christopher W. and Ochoa, Gabriela},
  title     = {Understanding Parameter Spaces Using Local Optima Networks: A Case Study on Particle Swarm Optimization},
  booktitle = {Proceedings of the 2021 Genetic and Evolutionary Computation Conference Companion},
  pages     = {1657--1664},
  publisher = {ACM},
  year      = {2021},
  doi       = {10.1145/3449726.3463145}
}

@inproceedings{TeixeiraPappa2022AutoMLLON,
  author    = {Teixeira, Matheus C. and Pappa, Gisele L.},
  title     = {Understanding {AutoML} Search Spaces with Local Optima Networks},
  booktitle = {Proceedings of the Genetic and Evolutionary Computation Conference},
  pages     = {449--457},
  publisher = {ACM},
  year      = {2022},
  doi       = {10.1145/3512290.3528743}
}

@incollection{MostertEtAl2019FeatureSelectionLON,
  author    = {Mostert, Werner and Malan, Katherine M. and Ochoa, Gabriela and Engelbrecht, Andries P.},
  title     = {Insights into the Feature Selection Problem Using Local Optima Networks},
  booktitle = {Evolutionary Computation in Combinatorial Optimization},
  editor    = {Liefooghe, Arnaud and Paquete, Lu{\'\i}s},
  series    = {Lecture Notes in Computer Science},
  volume    = {11452},
  pages     = {147--162},
  publisher = {Springer},
  address   = {Cham},
  year      = {2019},
  doi       = {10.1007/978-3-030-16711-0_10}
}

@article{ThomsonEtAl2020SamplingLON,
  author  = {Thomson, Sarah L. and Ochoa, Gabriela and Verel, S{\'e}bastien and Veerapen, Nadarajen},
  title   = {Inferring Future Landscapes: Sampling the Local Optima Level},
  journal = {Evolutionary Computation},
  volume  = {28},
  number  = {4},
  pages   = {621--641},
  year    = {2020},
  doi     = {10.1162/evco_a_00271}
}

@article{HomolyaVinko2019CentralityMDE,
  author  = {Homolya, Viktor and Vink{\'o}, Tam{\'a}s},
  title   = {Memetic Differential Evolution Using Network Centrality Measures},
  journal = {AIP Conference Proceedings},
  volume  = {2070},
  number  = {1},
  pages   = {020023},
  year    = {2019},
  doi     = {10.1063/1.5089990}
}

@incollection{HomolyaVinko2020LeveragingLONMDE,
  author    = {Homolya, Viktor and Vink{\'o}, Tam{\'a}s},
  title     = {Leveraging Local Optima Network Properties for Memetic Differential Evolution},
  booktitle = {Optimization of Complex Systems: Theory, Models, Algorithms and Applications},
  editor    = {Le Thi, Hoai An and Le, Hoai Minh and Pham Dinh, Tao},
  series    = {Advances in Intelligent Systems and Computing},
  volume    = {991},
  pages     = {109--118},
  publisher = {Springer},
  address   = {Cham},
  year      = {2020},
  doi       = {10.1007/978-3-030-21803-4_11}
}

@inproceedings{ThomsonEtAl2024MorphoLON,
  author    = {Thomson, Sarah L. and Le Goff, L{\'e}ni and Hart, Emma and Buchanan, Edgar},
  title     = {Understanding Fitness Landscapes in Morpho-Evolution via Local Optima Networks},
  booktitle = {Proceedings of the Genetic and Evolutionary Computation Conference},
  pages     = {114--123},
  publisher = {ACM},
  year      = {2024},
  doi       = {10.1145/3638529.3654059}
}

@article{BlackwellMacQueen1973,
  author  = {Blackwell, David and MacQueen, James B.},
  title   = {Ferguson distributions via {P}{\'o}lya urn schemes},
  journal = {The Annals of Statistics},
  volume  = {1},
  number  = {2},
  pages   = {353--355},
  year    = {1973}
}

@article{PitmanYor1997,
  author  = {Pitman, Jim and Yor, Marc},
  title   = {The two-parameter {P}oisson--{D}irichlet distribution derived from a stable subordinator},
  journal = {The Annals of Probability},
  volume  = {25},
  number  = {2},
  pages   = {855--900},
  year    = {1997}
}

@article{AlettiCrimaldi2021Twitter,
  author  = {Aletti, Giacomo and Crimaldi, Irene},
  title   = {Twitter as an innovation process with damping effect},
  journal = {Scientific Reports},
  volume  = {11},
  pages   = {21243},
  year    = {2021}
}

@article{AlettiEtAl2023,
  title={Interacting innovation processes},
  author={Aletti, Giacomo and Crimaldi, Irene and Ghiglietti, Andrea},
  journal={Scientific Reports},
  volume={13},
  number={1},
  pages={17187},
  year={2023},
  publisher={Nature Publishing Group UK London}
}

@article{AlettiEtAl2025,
  title={Central limit theorems for interacting innovation processes, related statistical tools and general results},
  author={Aletti, Giacomo and Crimaldi, Irene and Ghiglietti, Andrea},
  journal={arXiv preprint arXiv:2501.09648},
  year={2025}
}

@article{BellinaEtAl2025,
  title={Full spectrum of novelties in time-dependent urn models},
  author={Bellina, Alessandro and De Marzo, Giordano and Loreto, Vittorio},
  journal={Physical Review Research},
  volume={7},
  number={2},
  pages={023127},
  year={2025},
  publisher={APS}
}

@article{BarabasiBurstiness2005,
  author  = {Barab{\'a}si, Albert-L{\'a}szl{\'o}},
  title   = {The origin of bursts and heavy tails in human dynamics},
  journal = {Nature},
  volume  = {435},
  pages   = {207--211},
  year    = {2005}
}

@article{BaumannEtAl2019,
  author  = {Baumann, Oliver and Schmidt, Jan and Stieglitz, Nils},
  title   = {Effective search in rugged performance landscapes: A review and outlook},
  journal = {Journal of Management},
  volume  = {45},
  number  = {1},
  pages   = {285--318},
  year    = {2019}
}

@article{BillingerEtAl2014,
  author  = {Billinger, Stephan and Stieglitz, Nils and Schumacher, Terry R.},
  title   = {Search on rugged landscapes: An experimental study},
  journal = {Organization Science},
  volume  = {25},
  number  = {1},
  pages   = {93--108},
  year    = {2014}
}

@article{Frenken2006Modularity,
  author  = {Frenken, Koen},
  title   = {A fitness landscape approach to technological complexity, modularity, and vertical disintegration},
  journal = {Structural Change and Economic Dynamics},
  volume  = {17},
  number  = {3},
  pages   = {288--305},
  year    = {2006}
}

@article{GancoEtAl2020,
  author  = {Ganco, Martin and Kapoor, Rahul and Lee, Gwendolyn K.},
  title   = {From rugged landscapes to rugged ecosystems: Structure of interdependencies and firms' innovative search},
  journal = {Academy of Management Review},
  volume  = {45},
  number  = {3},
  pages   = {646--674},
  year    = {2020}
}

@article{IacopiniEtAl2020,
  author  = {Iacopini, Iacopo and Di Bona, Gabriele and Ubaldi, Enrico and Loreto, Vittorio and Latora, Vito},
  title   = {Interacting discovery processes on complex networks},
  journal = {Physical Review Letters},
  volume  = {125},
  pages   = {248301},
  year    = {2020}
}

@article{LiChenYing2019,
  title={Innovation search scope, technological complexity, and environmental turbulence: A NK simulation},
  author={Li, Fei and Chen, Jin and Ying, Ying},
  journal={Sustainability},
  volume={11},
  number={16},
  pages={4279},
  year={2019},
  publisher={MDPI}
}

@article{Price1965CitationNetworks,
  author  = {Price, Derek J. de Solla},
  title   = {Networks of Scientific Papers},
  journal = {Science},
  volume  = {149},
  number  = {3683},
  pages   = {510--515},
  year    = {1965},
  doi     = {10.1126/science.149.3683.510}
}

@article{FlemingSorenson2001Recombination,
  author  = {Fleming, Lee and Sorenson, Olav},
  title   = {Technology as a Complex Adaptive System: Evidence from Patent Data},
  journal = {Research Policy},
  volume  = {30},
  number  = {7},
  pages   = {1019--1039},
  year    = {2001},
  doi     = {10.1016/S0048-7333(00)00135-9}
}

@article{AharonsonSchilling2016TechLandscape,
  author  = {Aharonson, Barak S. and Schilling, Melissa A.},
  title   = {Mapping the Technological Landscape: Measuring Technology Distance, Technological Footprints, and Technology Evolution},
  journal = {Research Policy},
  volume  = {45},
  number  = {1},
  pages   = {81--96},
  year    = {2016},
  doi     = {10.1016/j.respol.2015.08.001}
}

\end{document}